\documentstyle[12pt,epsf]{article}
\begin{document}
\title{Large-Order Behavior of Two-coupling Constant
$\phi^4$-Theory with Cubic Anisotropy
}
\author{H.\ Kleinert and S.\ Thoms \\
Institut f\"ur Theoretische Physik \\
Freie Universit\"at Berlin \\
Arnimallee 14 \\
14195 Berlin}
\date{ }
\maketitle
\begin{abstract}
For the anisotropic $[u (\sum_{i=1}^N {\phi}_i^2)^2+
v \sum_{i=1}^N \phi_i^4]$-theory with \mbox{$N=2,3$} we calculate the
imaginary parts of the renormalization-group functions
in the form of a series expansion in $v$, \mbox{i.\ e.,}
around the isotropic case.
Dimensional regularization is used to evaluate the fluctuation
determinants for the isotropic instanton near the space dimension $4$.
The vertex functions in the presence of instantons are renormalized with the
help of a nonperturbative procedure introduced for
the simple $g{\phi}^4$-theory by McKane et al.
\end{abstract}
\section{Introduction}
More than twenty years ago, Brezin, Le Guillou and Zinn-Justin (BGZ)
studied the
phase transition of a cubic anisotropic system by means of renormalization
group equations \cite{BreGuZ1}.
Within a $(4-\varepsilon)$-expansion, they found that to lowest
nontrivial order in $\varepsilon$, the only stable
fixed point for $N < N_c = 4$ is
the ${\rm O}(N)$-symmetric one, where $N$ is the number of field components
appearing in the cubic anisotropic model.
They interpreted this as an indication
that the anisotropy is irrelevant as long as $N$ is smaller than four.
For $N > 4$, the isotropic fixed point
destabilized and the trajectories crossed over to the cubic fixed point.

Recently, our knowledge of perturbation coefficients
of the renormalization group functions of the anisotropic system was extended
up to the five-loop level by Kleinert and Schulte-Frohlinde \cite{KleSchu2}.
Since the perturbation expansions are badly divergent, they
do not directly yield improved estimates for the crossover
value $N_c$ where the isotropic fixed point destabilizes in favor of the
cubic one. An estimate using Pad\'{e} approximants \cite{KleSchu2}
indicates $N_c$ to lie below $3$, thus permitting real
crystals to exhibit critical exponents of the cubic universality class.

For a simple $\phi^4$-theory, the Pad\'{e} approximation is known to be
inaccurate.
In fact, the most accurate renormalization group functions for that theory
have been obtained by combining perturbation expansions
with large-order estimates and using a resummation procedure based on
Borel-transformations  \cite{KleJan}--\cite{GuZin5}.

It is the purpose of this paper to derive the large-order behavior of the
renormalization group functions for the anisotropic $g {\phi}^4$-theory.
In a forthcoming paper we will combine these results with the five-loop
perturbation expansion of Kleinert and Schulte-Frohlinde to derive the
precise value for the crossover value $N_c$.

For the simple $g {\phi}^4$-theory,
the large-order behavior of perturbation coefficients
has been derived by Lipatov \cite{Lip1,Lip2},
BGZ \cite{BreGuZ2} and others \cite{Bog}--\cite{Par2}
in a number of papers. The generalization to the ${\rm O}(N)$-symmetric case
was given in \cite{BreGuZ2}.
An equivalent method for calculating the large-order behavior
is based on the observation that
for a negative coupling constant Green functions possess an
exponentially small imaginary part due to the fact that the
ground state is unstable \cite{Lang,CollSop}.
The imaginary part is associated with the tunneling decay rate
of the ground state. It determines directly the large-order behavior of
the perturbation coefficients via a dispersion relation in the
complex coupling constant plane.

In the semiclassical limit,
the imaginary part of all Green functions can be calculated with the
help of classical solutions called {\em instantons}.
For a massless $g {\phi}^4$-theory in $d = 4$ space dimensions, these
instantons can be found analytically.
The imaginary part is a consequence of a negative frequency mode in the
spectrum of the fluctuation operator, whose determinant enters the
one-loop correction to the instanton contribution.
McKane, Wallace and de Alcantara Bonfim \cite{WalBon} found a way to
continue the results of the
$g {\phi}^4$-theory in $d = 4$ to a field theory in $d = 4-\varepsilon$
dimensions. They proposed an extended dimensional regularization scheme
for nonperturbative renormalizing the imaginary parts of vertex functions.

In the present work we have to extend this scheme to the case of
a ${\phi}^4$-theory with cubic anisotropy, where the energy functional
has the following form:
\begin{equation}
\label{HN}
H(\vec{\phi})=\displaystyle{\int}d^dx
\left[\frac{1}{2} \sum_{i=1}^N {\phi}_i (-{\nabla}^2) {\phi}_i +
\frac{u}{4} (\sum_{i=1}^N {\phi}_i^2)^2+
\frac{v}{4} \sum_{i=1}^N \phi_i^4 \right] \,\, .
\end{equation}
For $N = 2$, the corresponding model in quantum mechanics
was first studied by Banks, Bender and Wu \cite{BaBeWu,BaBe}
who used multidimensional WKB-techniques
to derive the large-order behavior of the perturbation
expansion for the ground state energy.
In $1990$ Janke \cite{Jan} presented a more efficient calculation
using a path integral approach. In the present work, this approach will be
generalized to quantum field theory and extended by a careful discussion
of the region near the isotropic limit $v \rightarrow 0$.
This is important, since the infrared-stable cubic fixed point
is expected to appear very close to the ${\rm O}(N)$-symmetric one.
In fact, it will be sufficient to give the quantum-field
theoretical generalization of \cite{Jan} in terms of an expansion
about the isotropic case in powers of $v$.

The paper is organized as follows.
The method is developed by treating first the case $N=2$.
In \mbox{Section $2$} we derive the Feynman rules
for the power series expansion of all Green functions
around the isotropic limit. In \mbox{Section $3$} we calculate
the small-oscillation determinants for the transversal and longitudinal
fluctuations. In \mbox{Section $4$} we use the
extended renormalization scheme of \cite{WalBon} to find the
full (real and imaginary) vertex functions, and derive the renormalization
constants to one loop. In \mbox{Section $5$} we calculate
the imaginary parts of the renormalization-group functions and thus the
large-order behavior of the perturbation coefficients.
In \mbox{Section $6$}, finally, we extend the results to the
physically relevant case $N=3$.
\section{Fluctuations around the isotropic instanton}
For positive coupling constants $u$ and $v$, the
system defined by the energy functional (\ref{HN}) is
stable and the Green functions are
real.
On the other hand, if both coupling constants are negative
the system is unstable and the Green functions acquire an imaginary part.
The corresponding functional integrals can be calculated by an analytical
continuation from positive to negative coupling constants, keeping the factor
$\exp{[\int (-\frac{u}{4}
(\sum_{i=1}^N \phi_i^2)^2 -\frac{v}{4}
\sum_{i=1}^N \phi_i^4)\,d^4x]}$
real.
We perform this analytical continuation by means of a joint rotation in the
two complex planes, substituting
$u \rightarrow u \exp{(i\theta)}$ and
$v \rightarrow v \exp{(i\theta)}$,
and rotating the azimuthal angle $\theta$
from $0$ to $\pi$. At the same time, we rotate the
contour of integration in the field space. The convergence of the
functional integrals is maintained by the field rotation
$\phi \rightarrow \phi \exp{(-i{\theta}/4)}$.

A natural parameter for the anisotropy of the system is the ratio
$\delta = v /(u+v)$. The isotropic limit corresponds to $\delta = 0$.
During the joint rotation of $u$ and
$v$, the parameter $\delta$ remains constant.
Thus, $\delta$ is a good parameter for the anisotropy at both
positive and negative couplings $u$ and $v$.
We shall use the coupling constants
$g=u+v$ and $\delta$ for a calculation of the
Green functions. These are given by the functional integrals
\begin{eqnarray}
\label{G2M}
\lefteqn{G^{(2M)} (x_1,x_2,\ldots,x_{2M})_{i_1,i_2,\ldots,i_{2M}} = }
\nonumber \\
\nonumber \\
    & &\frac{\int D \phi \,\,
       \phi_{i_1} (x_1) \phi_{i_2} (x_2) \cdot \ldots \cdot
       \phi_{i_{2M}} (x_{2M})
       \exp\{-H[\phi]\} }{\int D \phi \exp\{
       -H[\phi]\} } \, ,
\nonumber \\
\end{eqnarray}
where the subscripts $i_k$ run through the $N$ components of the field
$\phi$. As explained in the introduction, we shall first study the case $N=2$
with the free energy functional
\begin{eqnarray}
\label{Hn2}
\lefteqn{H[\phi_1,\phi_2] = }
\nonumber \\
\nonumber \\
    & &\!\!\!\!\!\!\!\!
       \int d^dx \left\{\frac{1}{2} \left[\phi_1 (-{\nabla}^2) \phi_1+
       \phi_2 (-{\nabla}^2) \phi_2 \right]+\frac{g}{4}  \left[\phi_1^4+
       2(1-\delta) \phi_1^2 \phi_2^2 +
       \phi_2^4 \right] \right\}.
\end{eqnarray}

When expressed in terms of the coupling constants $g$ and $\delta$, the
Green functions possess an imaginary part for $g < 0$.
For the reasons explained above, we shall derive an expansion of
the imaginary parts of the Green functions (\ref{G2M})
around the isotropic case,
i.~e., in powers of $\delta$:
\begin{equation}
\label{IMG}
{\rm Im} \, G =
\sum_{n=0}^{\infty} a_n \, {\delta}^n \left(\frac{A}{-g}\right)^{p(n)}
\exp \left(\frac{A}{g}\right) [1+{\cal O}(g)]\, .
\end{equation}
Each power ${\delta}^n$ has its own $n$-dependent imaginary
part.
Given such an expansion, the large-order estimates for the coefficients of the
powers $g^k$ follows from a dispersion relation in $g$:
\begin{equation}
\label{DR}
G=\frac{1}{\pi}\int\limits_{-\infty}^0 d\bar{g}\,
\frac{{\rm Im}\, G(\bar{g}+i0)}{\bar{g}-g}\, .
\end{equation}
For a general discussion of the relationship between imaginary parts
and large-order behavior see, for example, ch.\ $17$ of \cite{book}.
If the power-series expansion of $G$ is
\begin{equation}
\label{LO1}
G=\sum_{k,n=0}^{\infty}G_{kn}\, g^k {\delta}^n \, ,
\end{equation}
the coefficients $G_{kn}$ have the asymptotic behavior
\begin{equation}
\label{LO2}
G_{kn}\stackrel{k \rightarrow \infty}{\longrightarrow}\,
-\frac{a_n}{\pi}\left(-1\right)^k \left(\frac{1}{A}\right)^k k!\, k^{p(n)-1}
\left[1+{\cal O} \left(1/k \right)\right].
\end{equation}
These estimates apply to perturbation coefficients,
in which the maximal power of $g$ is much larger than the power of $\delta$.
Being interested in the region close to the isotropic limit,
this restricted large-order estimation will be sufficient.

An expansion of the Green function (\ref{G2M}) around the instanton yields
exponentially small imaginary parts in both numerator and denominator.
In order to isolate the imaginary part of the numerator, we simplify
the denominator in (\ref{G2M}) and calculate first the imaginary part
of the approximate Green functions
\begin{eqnarray}
\label{TilG}
\lefteqn{\tilde{G}^{(2M)} (x_1,x_2,\ldots,x_{2M})_{i_1,i_2,\ldots,i_{2M}} = }
\nonumber \\
\nonumber \\
\!\!\!\!\!& &\frac{\int D \phi \,\,
       \phi_{i_1} (x_1) \phi_{i_2} (x_2) \cdot \ldots \cdot
       \phi_{i_{2M}} (x_{2M})
       \exp\{-H[\phi]\} }{\int D \phi \exp\{
       -H_0 [\phi]\} }\, ,
\nonumber \\
\end{eqnarray}
where the denominator contains only the free energy functional $H_0$.

With the aim of calculating (\ref{IMG}),
we expand the fields around the classical solution of the
isotropic limit $\delta = 0$ for the space dimension $d = 4$. Thus we write:
\begin{equation}
\label{exp}
\vec{\phi}=\vec{u}_L \phi_c+\vec{u}_L \xi+\vec{u}_T \eta=
\left( \begin{array}{r}
       \cos{\vartheta} \\ \sin{\vartheta}
       \end{array} \right) \phi_c+
\left( \begin{array}{r}
       \cos{\vartheta} \\ \sin{\vartheta}
       \end{array} \right) \xi+
\left( \begin{array}{r}
       -\sin{\vartheta} \\ \cos{\vartheta}
       \end{array} \right) \eta \, ,
\end{equation}
where $\phi_c$ is the well-known $g {\phi}^4$-instanton in four dimensions
\begin{equation}
\label{fcl}
\phi_c=\left(\frac{8}{-g}\right)^{1/2}
\frac{\lambda}{1+{\lambda}^2(x-x_0)^2} \, ,
\end{equation}
and $\vartheta$ is the rotation angle of the isotropic instanton in the
$(\phi_1,\phi_2)$-plane. The fields $\xi$ and $\eta$ correspond to the degrees
of freedom orthogonal to the rotation of that instanton.
The parameters $x_0$ and $\lambda$ are position
and scale size of the instanton, respectively. Inserting the
expansion (\ref{exp}) in $H(\phi_1,\phi_2)$, we obtain the expression:
\begin{equation}
\label{Hall}
H(\phi_1,\phi_2)=H(\phi_{1c},\phi_{2c})+H_1+H_2+H_3+H_4 \, ,
\end{equation}
where $H_i$ collects all terms in $\xi$ and $\eta$ of order $i$.
They are given by: \\
(1) Linear terms:
\begin{eqnarray}
\label{H1}
H_1(\xi,\eta)
      &=&-\frac{\varepsilon 4(2)^{1/2}}{(-g)^{1/2}}
         \int d^dx\frac{\lambda^3}{[1+\lambda^2(x-x_0)^2]^2} \xi
         \nonumber \\
      & &+\frac{\delta}{(-g)^{1/2}}\, \sin^2(2\vartheta)\,8(2)^{1/2}
         \int d^dx \frac{\lambda^3}{[1+\lambda^2(x-x_0)^2]^3} \xi
         \nonumber \\
      & &+\frac{\delta}{(-g)^{1/2}}\, \sin(4\vartheta) \,4(2)^{1/2}
         \int d^dx \frac{\lambda^3}{[1+\lambda^2(x-x_0)^2]^3} \eta
\end{eqnarray}
(2) Quadratic terms:
\begin{eqnarray}
\label{H2}
H_2(\xi,\eta)
      &=&\frac{1}{2} \int d^dx \xi M_L \xi+
         \frac{1}{2} \int d^dx \eta M_T \eta
         \nonumber \\
      & &+\delta \, \sin^2(2\vartheta) \, 6 \int d^dx
         \frac{\lambda^2}{[1+\lambda^2(x-x_0)^2]^2}(\xi^2-\eta^2)
         \nonumber \\
      & &+\delta \, \sin(4\vartheta)\, 6 \int d^dx
          \frac{\lambda^2}{[1+\lambda^2(x-x_0)^2]^2} \xi \eta \, ,
\end{eqnarray}
\hspace*{3ex} where $M_L$ and $M_T$ are the operators
\begin{equation}
\label{MLT}
M_L  = -{\nabla}^2-\frac{24 \lambda^2}{[1+\lambda^2 (x-x_0)^2]^2}\, , \,\,
M_T  = -{\nabla}^2-
\frac{ 8(1-\delta) \lambda^2}{[1+\lambda^2 (x-x_0)^2]^2}
\end{equation}
(3) Cubic terms:
\begin{eqnarray}
\label{H3}
H_3(\xi,\eta)
       &=&-(-8g)^{1/2} \int d^dx \frac{\lambda}{1+\lambda^2(x-x_0)^2}
          \left[\xi^3+(1-\delta) \xi \eta^2 \right]
          \nonumber \\
       & &-\delta \, \frac{\sin^2(2\vartheta)}{2} \, (-8g)^{1/2}
          \int d^dx \frac{\lambda}{1+\lambda^2(x-x_0)^2}
          (3 \eta^2 \xi-\xi^3)
          \nonumber \\
       & &-\delta \, \frac{\sin(4\vartheta)}{4} \, (-8g)^{1/2}
          \int d^dx \frac{\lambda}{1+\lambda^2(x-x_0)^2}
          (\eta^3-3 \eta \xi^2)
\end{eqnarray}
(4) Quartic terms:
\begin{eqnarray}
\label{H4}
H_4(\xi,\eta)
       &=& +\frac{g}{4}\int d^dx \left[\xi^4+
           \eta^4+2(1-\delta) \xi^2 \eta^2 \right]
           \nonumber \\
       & &-\delta \, g \, \frac{\sin^2(2\vartheta)}{8} \int d^dx
           (\xi^4+\eta^4-6\xi^2 \eta^2)
           \nonumber \\
       & &+\delta \, g \, \frac{\sin(4\vartheta)}{4} \int d^dx
           (\xi \eta^3-\eta \xi^3)
\end{eqnarray}
The linear terms (\ref{H1}) contain factors $\varepsilon$ or $\delta$,
since the expansion of the fields around the isotropic instanton is
extremal only in the four-dimensional isotropic limit.

The classical contribution of the instanton is
\begin{eqnarray}
\label{Hclas}
\lefteqn{H(\phi_{1c},\phi_{2c}) = }
\\
\nonumber \\
      & &-\frac{\lambda^{\varepsilon}}{g} \frac{8\pi^2}{3}
      \left[1-\frac{\varepsilon}{2}\left(2+\ln \pi+\gamma \right) \right]
      -\frac{\lambda^{\varepsilon} \delta }{g} \, \frac{8\pi^2}{3}
      \frac{\sin^2(2\vartheta)}{2}+
      {\cal O} \left(\frac{ \delta }{g}\, \varepsilon
      \right) \! .
\nonumber
\end{eqnarray}
The first term in $H(\phi_{1c},\phi_{2c})$ is evaluated up to the first
order in $\varepsilon$, because the one-loop renormalization will require
replacing $1/g \rightarrow 1/g_{\rm r}
+ {\cal O}[f({\delta}_{\rm r})/{\varepsilon}]$,
where $g_r$ is the renormalized version of the coupling constant $g$.
A contribution of order $\varepsilon$ is not needed in the second,
$\delta$-dependent term, where it would produce a further factor
$\delta$, and thus be part of the neglected terms
${\cal O}(g)$ in (\ref{IMG}).

Due to the anisotropy of the action, the fluctuation expansion
(\ref{H1})--(\ref{Hclas})
contains $\vartheta$-dependent parts. However, all these terms are of order
$\delta$ and can therefore be handled by straightforward perturbation theory.

Note that the angle $\vartheta$ disappears when the isotropic instanton is
directed along the coordinate axis. Then the isotropic instanton
coincides with the exact solution in $d = 4$ for $\delta > 0$. This is
also seen by inspecting the potential in (\ref{Hn2}).
For $\delta > 0$, the term ${\phi}_1^4 + {\phi}_2^4$ is larger than
$2 (1-\delta) {\phi}_1^2 {\phi}_2^2$,
so that the ``tunneling-paths'' of
extremal action are obviously straight lines along the coordinate axis.

For the region $\delta < 0$, the exact instanton follows from the known fact
that the action (\ref{Hn2})
is invariant under the orthogonal transformation:
$${\phi}_1 =({\tilde{\phi_1}} + {\tilde{\phi_2}})/{\sqrt{2}} \, , \qquad
{\phi}_2 =({\tilde{\phi_1}} - {\tilde{\phi_2}})/{\sqrt{2}} \, ,$$
with the new coupling constants:
$$g=(2-{\tilde{\delta}}) {\tilde{g}}/2 \, , \qquad
\delta = -\frac{2 \, \tilde{\delta}}{2-\tilde{\delta}} \, ,$$
satisfying $\delta < 0$ for $\tilde{\delta} > 0$.

In contrast to the method in \cite{Jan}, the treatment of the fluctuations
in a power series in $\delta$ does not allow us to deal with all modes
perturbatively.
Near the ${\rm O}(2)$-symmetric case, the Gaussian approximation
for the rotation of the instanton becomes invalid, and the Gaussian integral
must be replaced by an exact angle integration.
The separation of the rotation mode must be done with the help
of collective coordinates.
The Jacobian of the relevant transformation can be deduced from the
isotropic system.
The field $\eta$ of small oscillations must not contain modes of
$M_T$ with eigenvalues of the order $\delta$, since these would vanish for
$\delta = 0$. The discussion of the
longitudinal part is given in \cite{Wal}.

In dimensional regularization, all eigenvalues of $M_L$ which would be zero
for $d = 4$ are of order $\varepsilon$ in $4-\varepsilon$ dimensions. To avoid
eigenvalues of the order $1/{\varepsilon}$ in the
propagator for the longitudinal fluctuations, all collective
coordinates of the four-dimensional case must be retained in
$d$ dimensions. Therefore the field $\xi$ in (\ref{exp})
contains no modes of $M_L$ with eigenvalues proportional to $\varepsilon$.

In order to calculate the fluctuation factor and to separate the
almost-zero modes from $\det M_L$ and $\det M_T$, it is convenient to do a
conformal transformation onto a sphere in $d+1$ dimensions leading to
a discrete spectrum for the transformed
differential operators $M_L$ and $M_T$ \cite{Drum,DrumSho}.
Their products of eigenvalues can be given in terms of the Riemann
$\zeta$-function which we easily expanded near $\varepsilon=0$.
Simple $1/{\varepsilon}$-poles, which are characteristic
of dimensional regularization, appear directly from the known singularity
of that function. In place of the fields $\xi$ and $\eta$, we define the fields
${\Phi}_1 (\rho)$ and ${\Phi}_2 (\rho)$ on the unit sphere in $d+1$
dimensions:
\begin{equation}
\Phi_1={\sigma}^{1-d/2} \xi \, \quad \Phi_2={\sigma}^{1-d/2} \eta \, ,
\end{equation}
where $\sigma = 2/(1+x^2)$.
The instanton is supposed to be centered at the origin and
rescaled by $\lambda$. The spherical operator corresponding to the
differential operator ${\nabla}^2$ is
\begin{equation}
V_0=\frac{1}{2} L^2- \frac{1}{4} d(d-2)\, ,
\end{equation}
where $L^2 =\sum_{a,b} ({\rho}_a {\partial}_b-{\rho}_b {\partial}_a)^2$ is
the total angular momentum operator on the $(d+1)$-dimensional sphere.

The eigenfunctions of $V_0$ are the spherical
harmonics $Y_m^l(\rho)$ in $d+1$ dimensions \cite{Harm}.
They satisfy
\begin{equation}
V_0\, Y^l_m (\rho)=-(l+\frac{1}{2} d-1)(l+\frac{1}{2} d)\, Y^l_m(\rho)\, ,
\end{equation}
and have the degeneracy
\begin{equation}
\nu_l(d+1)=\frac{(2l+d-1) \Gamma(l+d-1)}{\Gamma(d) \Gamma(l+1)} \, .
\end{equation}
After the transformation,
the angle-independent quadratic part of $H_2$ in (\ref{H2}) becomes
\begin{equation}
\frac{1}{2} \int d \Omega \Phi_1 (-V_0-6) \Phi_1 +
\frac{1}{2} \int d \Omega \Phi_2 \left[-V_0-2 (1-\delta) \right] \Phi_2 \, ,
\end{equation}
where $d{\Omega}$ is the surface element of the unit sphere in $d+1$
dimensions.

After rewriting the entire energy functional(\ref{Hall}) in terms of the new
fields,
we can summarize the Feynman rules for the diagrammatic evaluation of the
functional integrals as follows: \\
$(1)$ Propagators (from the $\vartheta$-independent part of $H_2$):
\\
\\
\hspace*{2em}~longitudinal propagator:
\begin{equation}
\epsfxsize=2cm
\epsfbox{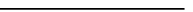}\,\, =\, \left[\left(-V_0-6\right)^{'}\right]^{-1}
\end{equation}
\hspace*{2em}~transversal propagator:
\begin{equation}
\hspace*{6ex}
\epsfxsize=2cm
\epsfbox{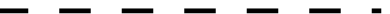}\, \, =\, \left\{\left[-V_0-2(1-\delta)
\right]^{'}\right\}^{-1}
\end{equation}
The prime indicates that the almost-zero modes are omitted when forming
the inverse.\\
$(2)$ Tadpoles $(H_1)$:
\begin{eqnarray}
\epsfxsize=1.5cm
\epsfbox{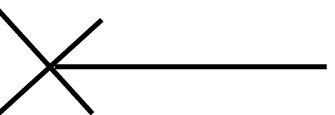} &\raisebox{1ex}{=}&\!\!
\raisebox{1ex}{$\displaystyle{
-\varepsilon \left(\frac{2}{-g}\right)^{1/2}
\sigma^{-1+{\varepsilon}/2} \Phi_1}$}
\nonumber \\
\nonumber \\
\epsfxsize=1.5cm
\epsfbox{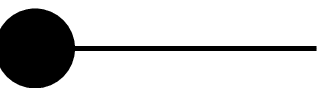}
&\raisebox{1ex}{=}&\!\!
\raisebox{1ex}{$\displaystyle{
\left(\frac{2}{-g}\right)^{1/2} \delta \, \sin^2(2\vartheta)
\sigma^{{\varepsilon}/2} \Phi_1}$}
\nonumber \\
\nonumber \\
\epsfxsize=1.5cm
\epsfbox{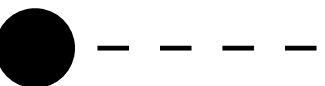} &\raisebox{1ex}{=}&\!\!
\raisebox{1ex}{$\displaystyle{
\left(\frac{1}{-2g}\right)^{1/2} \delta \, \sin(4\vartheta)
\sigma^{{\varepsilon}/2} \Phi_2}$}
\nonumber \\
\end{eqnarray}
The bold dot stands for the $\vartheta$-dependence of the vertex. \\
$(3)$ Two-point vertices ($\vartheta$-dependent part of $H_2$): \\
\begin{eqnarray}
\epsfxsize=2.5cm
\epsfbox{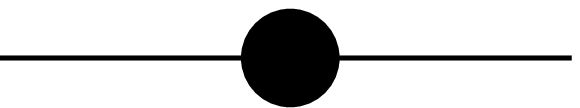}\,\, \raisebox{1ex}{+} \,\,
\epsfxsize=2.5cm
\epsfbox{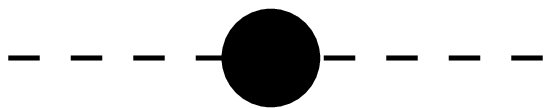} &\raisebox{1ex}{=}&\!\!
\raisebox{1ex}{$\displaystyle{
\frac{3}{2}\sin^2(2\vartheta) \, \delta \, \left(\Phi_1^2-\Phi_2^2 \right)}$}
\nonumber \\
\nonumber \\
\epsfxsize=2.5cm
\epsfbox{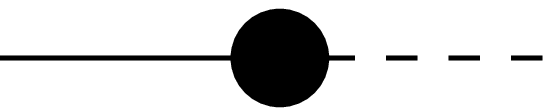} &\raisebox{1ex}{=}&\!\!
\raisebox{1ex}{$\displaystyle{
\frac{3}{2} \sin(4\vartheta) \, \delta \, \Phi_1 \Phi_2}$}
\nonumber \\
\end{eqnarray}
$(4)$ Cubic vertices $(H_3)$:
\begin{eqnarray}
\epsfxsize=1cm
\epsfbox{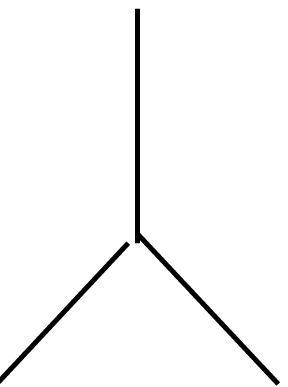} &\raisebox{3ex}{=}& \!\!
\raisebox{3ex}{$\displaystyle{
-(-2g)^{1/2}\sigma^{-{\varepsilon}/2} \Phi_1^3}$} \nonumber \\
\epsfxsize=1cm
\epsfbox{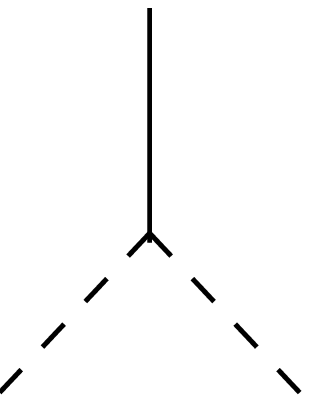} &\raisebox{3ex}{=}&\!\!
\raisebox{3ex}{$\displaystyle{
-(-2g)^{1/2} (1-\delta)\sigma^{-{\varepsilon}/2}
\Phi_1 \Phi_2^2}$} \nonumber \\
\epsfxsize=1cm
\epsfbox{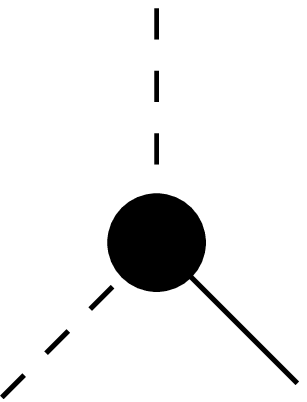}\,\, \raisebox{3ex}{+} \,\,
\epsfxsize=1cm
\epsfbox{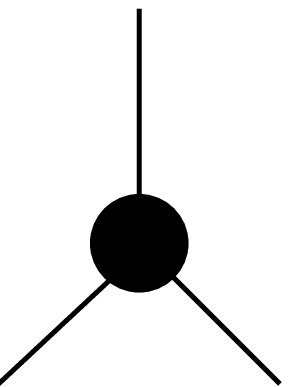} &\raisebox{3ex}{=}& \!\!
\raisebox{3ex}{$\displaystyle{
-(-2g)^{1/2} \, \delta \, \frac{\sin^2(2\vartheta)}{2}
\sigma^{-{\varepsilon}/2}\left(3\Phi_2^2 \Phi_1-\Phi_1^3 \right)}$}
\nonumber \\
\epsfxsize=1cm
\epsfbox{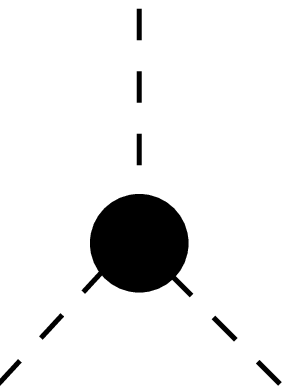}\,\, \raisebox{3ex}{+} \,\,
\epsfxsize=1cm
\epsfbox{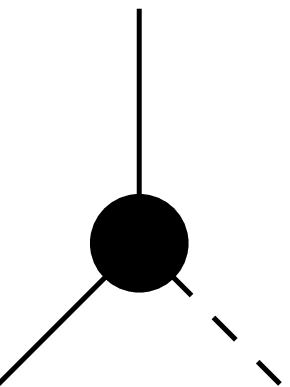} &\raisebox{3ex}{=}&\!\!
\raisebox{3ex}{$\displaystyle{
-(-2g)^{1/2} \, \delta \, \frac{\sin(4\vartheta)}{4}
\sigma^{-{\varepsilon}/2} \left(\Phi_2^3-3\Phi_2 \Phi_1^2 \right)}$}
\nonumber \\
\end{eqnarray}
$(5)$ Quartic vertices $(H_4)$:
\begin{eqnarray}
\epsfxsize=1cm
\epsfbox{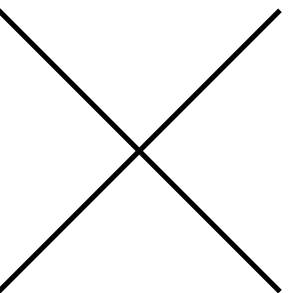}\,\, \raisebox{2.5ex}{+} \,\,
\epsfxsize=1cm
\epsfbox{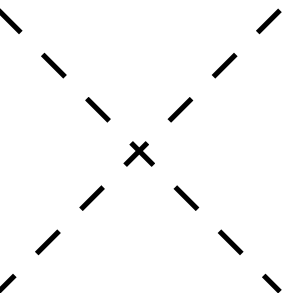}\,\, \raisebox{2.5ex}{+} \,\,
\epsfxsize=1cm
\epsfbox{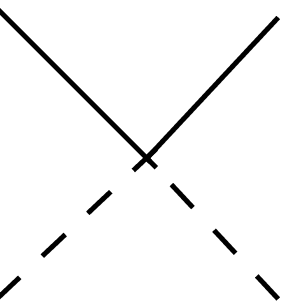} &\raisebox{2.5ex}{=}& \!\!
\raisebox{2.5ex}{$\displaystyle{
\frac{g}{4} \sigma^{-\varepsilon} \left(\Phi_1^4+\Phi_2^4+
2(1-\delta) \Phi_1^2 \Phi_2^2 \right)}$}
\nonumber \\
\nonumber \\
\epsfxsize=1cm
\epsfbox{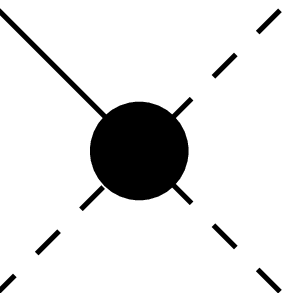}\,\, \raisebox{2.5ex}{+} \,\,
\epsfxsize=1cm
\epsfbox{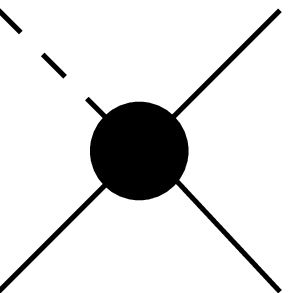} &\raisebox{2.5ex}{=}& \!\!
\raisebox{2.5ex}{$\displaystyle{
\frac{g}{4}\, \delta \,\sin(4\vartheta)
\sigma^{-\varepsilon} \left(\Phi_1 \Phi_2^3-
\Phi_2 \Phi_1^3 \right)}$}
\nonumber \\
\nonumber \\
\epsfxsize=1cm
\epsfbox{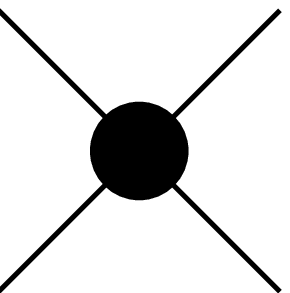}\,\, \raisebox{2.5ex}{+} \,\,
\epsfxsize=1cm
\epsfbox{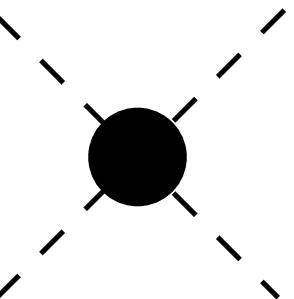}\,\, \raisebox{2.5ex}{+} \,\,
\epsfxsize=1cm
\epsfbox{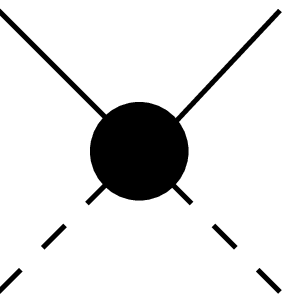} &\raisebox{2.5ex}{=}&\!\!
\raisebox{2.5ex}{$\displaystyle{
-\frac{g}{4} \, \delta \, \frac{\sin^2(2\vartheta)}{2}
\sigma^{-\varepsilon} \left(\Phi_1^4 +\Phi_2^4-6\Phi_1^2 \Phi_2^2 \right)}$}
\nonumber \\
\end{eqnarray}
In order to obtain the leading
corrections we must consider all connected diagrams of order ${\cal O}(1/g)$,
${\cal O}({\delta}/g)$, and ${\cal O}(1)$.
The contributions of these diagrams are
added to $H_c$ in (\ref{Hclas}). We consider
first the contributions to ${\cal O}(1/g)$, which
arise from the connected tree diagrams
\begin{equation}
\epsfxsize=2cm \epsfbox{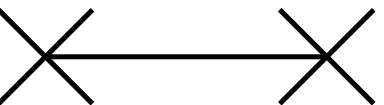}\, , \qquad
\epsfxsize=1.5cm \epsfbox{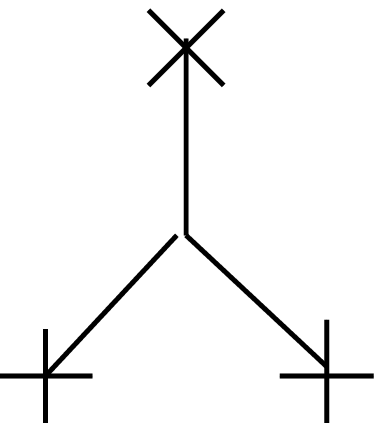}\, , \qquad
\epsfxsize=1.5cm \epsfbox{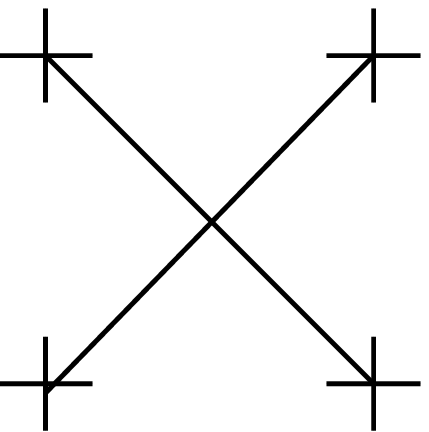}
\end{equation}
and their ${\Phi}_1$ insertions of zeroth order in $g$:
\begin{equation}
\epsfxsize=1.5cm \epsfbox{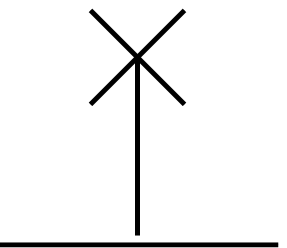}\, ,\quad
\epsfxsize=2cm \epsfbox{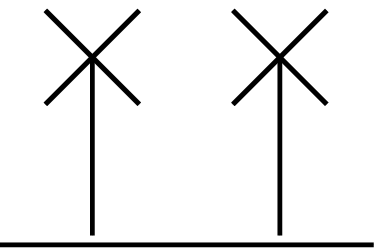}\, , \quad
\epsfxsize=2.5cm \epsfbox{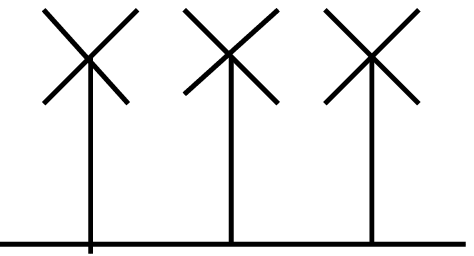}\, \quad \ldots
\end{equation}
Since each ${\Phi}_1$-vertex produces an $\varepsilon$-factor, the smallest
order in $\varepsilon$ is two. However, terms of order ${\varepsilon}^2/g$
are negligible for our calculation even after one-loop renormalization, since:
$1/g \rightarrow 1/g_{\rm r} + {\cal O}[f({\delta}_{\rm r})/{\varepsilon}]$ and
$\varepsilon \rightarrow 0$. The diagrams of ${\cal O}({\delta}/g)$ can be
generated from those of ${\cal O}(1/g)$ by one of the substitutions:
\begin{equation}
\epsfysize=1cm \epsfbox{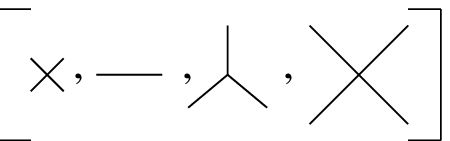} \quad
\raisebox{2ex} {$\Longrightarrow$}
\quad \epsfysize=1cm \epsfbox{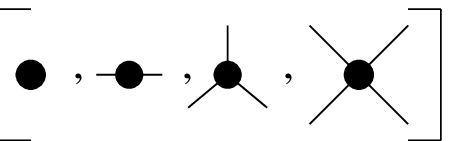} \raisebox{2ex}{$\, \, .$}
\end{equation}
By inspection, we see that the only diagram with an $\varepsilon$-power
smaller than two is given by
\begin{equation}
\epsfxsize=2cm \epsfbox{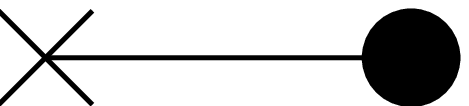}
\end{equation}
which is of order $\varepsilon \, {\delta}/g$.
After the renormalization, a term of order ${\delta}_{\rm r}$ appears.
But this term can be neglected in comparison
with the ${\delta}_{\rm r}/g_{\rm r}$-term
from $H_c$ in expression (\ref{Hclas}).
Hence all the tree diagrams would enter only in a higher-order calculation.

Contributions to order $g^0$ can appear
from one-loop diagrams. The only possible
candidates are
\begin{equation}
\epsfxsize=7cm \epsfbox{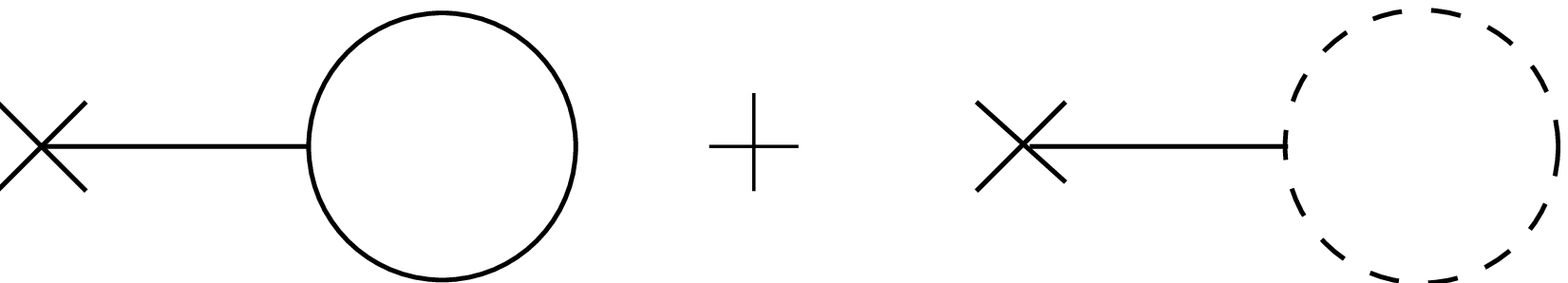} \raisebox{3ex}{$\, \, ,$}
\end{equation}
where the $1/{\varepsilon}$-pole from the loop integration and
the $\varepsilon$-factor of the ${\Phi}_1$-vertex cancel.
Following the method of \cite{Wal}, we have found that these diagrams
contribute to the coefficient of the imaginary part of $\tilde{G}^{(2M)}$
in (\ref{TilG}) a factor
\begin{equation}
c_1^L c_1^T \, =\exp(-5+\delta-{\delta}^2/2) \, .
\end{equation}
\section{Quadratic Fluctuation Determinants}
The angle-independent quadratic form is decomposed into a longitudinal and a
transverse part. For the longitudinal fluctuations we obtain the
determinant
\begin{equation}
\label{Lprod}
\left(\frac{\det V_L}{\det V_{0L}} \right)^{-\frac{1}{2}} \!\!=
\left[\frac{\det(-V_0-6)}{\det(-V_0)} \right]^{-\frac{1}{2}} \!\!=
\prod^{\infty}_{l=0} \left[\frac{(l+\frac{1}{2} d-3)(l+\frac{1}{2} d+2)}{
(l+\frac{1}{2} d-1)(l+\frac{1}{2} d)} \right]^{-\frac{1}{2} \nu_l(d+1)}_{,}
\end{equation}
which coincides with that appearing in the one-component $g{\phi}^4$ theory,
and is therefore known \cite{Wal}. It contains
a bound state at $l=0$, this being
responsible for the expected imaginary part of $\tilde{G}^{(2M)}$, and $d+1$
almost-zero eigenmodes of order $\varepsilon$ associated with the
dilatation and translation degrees of freedom of the instanton in
$4-\varepsilon$ dimensions. Extracting these modes from the
product (\ref{Lprod})
in the framework of collective coordinates \cite{ZittLan,GeSa},
we obtain the well-known formal
replacement rule for the determinant
\begin{equation}
\left(\frac{\det V_L}{\det V_{0L}} \right)^{-\frac{1}{2}} \Longrightarrow
J^{V_L} (2\pi)^{-\frac{(d+1)}{2}} c_2^L
\end{equation}
with
\begin{equation}
c_2^L=(2\pi)^{-1/2} 5^{5/2} \exp \left[ \frac{3}{\varepsilon}+\frac{3}{4}-
\frac{7}{2} \gamma+\frac{3}{\pi^2} \zeta^{'}(2) \right] \, ,
\end{equation}
and the Jacobian of the collective coordinates transformation:
\begin{equation}
J^{V_L}=\lambda^{(d-1)} \left(-\frac{16\pi^2 \lambda^{\varepsilon}}{
15g} \right)^{\frac{d+1}{2}} [1+{\cal O}(\varepsilon,g)] \, .
\end{equation}
The expression for $c_2^L$ contains the Euler constant $\gamma$, and the
derivative of the Riemann-zeta function $\zeta(x)$ at the value
$x=2$. Note the simple pole in $\varepsilon$, which is related to the
ultraviolet divergence on the one-loop level.

The transverse fluctuations
contain neither negative nor zero modes for $\delta$ larger than zero.
The corresponding fluctuation determinant is given by
\begin{eqnarray}
\label{Tprod}
\left(\frac{\det V_T}{\det V_{0T}}
\right)^{-\frac{1}{2}} &=& \left\{\frac{\det \left[-V_0-2(1-\delta) \right]}
                           {\det(-V_0)} \right\}^{-\frac{1}{2}}
\nonumber \\
\nonumber \\
                       &=& \prod^{\infty}_{l=0} \left[\frac{(l+
                       \frac{1}{2} d-1)(l+\frac{1}{2} d)-2(1-\delta)}
                       {(l+\frac{1}{2} d-1)(l+\frac{1}{2} d)}
                       \right]^{-\frac{1}{2} \nu_l(d+1)}.
\end{eqnarray}
Just one zero eigenmode appears for $l=0$ in the isotropic limit
$\delta \rightarrow 0$, due to the rotational invariance in that case. For
$\delta > 0$, the numerator in (\ref{Tprod}) contributes a factor
$1/{\sqrt{2 \delta}}$. To avoid this artificial
zero-mode divergence for $\delta \rightarrow 0$ we separate,
as announced earlier, the angular
degree of freedom from the integral measure via the collective-coordinates
method. Similar to the longitudinal case, we make the
formal substitution
\begin{equation}
\left(\frac{\det V_T}{\det V_{0T}} \right)^{-\frac{1}{2}} \Longrightarrow
J^{V_T} (2\pi)^{-\frac{1}{2}} c_2^T
\end{equation}
with
\begin{equation}
J^{V_T}=\left(-\frac{16\pi^2 \lambda^{\varepsilon}}{
3g} \right)^{\frac{1}{2}} \, ,
\end{equation}
which coincides with the Jacobian of the isotropic model.
Since the integration interval for the angle $\vartheta$ is compact,
no singularity appears in the limit $\delta \rightarrow 0$.

It is useful to illustrate the appearance of a divergent factor
$1/{\sqrt{2 \delta}}$ in a careless use of Gaussian integral.
Assuming $\delta > 0$, we expand the angle-dependent
classical action around $\vartheta = 0$ up to quadratic order.
The ensuing Gaussian integral can be evaluated after replacing the finite
integration interval $\vartheta \in [-\pi,\pi]$ by the noncompact one
$\vartheta \in [-\infty,\infty]$.
In this way, the integral
\begin{displaymath}
J^{V_T}(2\pi)^{-\frac{1}{2}}
\int\limits_{-\pi}^{+\pi} d \vartheta \exp
\left[-\lambda^{\varepsilon} \frac{8\pi^2}{3} \, \frac{\delta}{|g|} \,
\frac{\sin^2(2\vartheta)}{2} \right]
\end{displaymath}
is evaluated to
\begin{displaymath}
\left(\frac{8\pi \lambda^{\varepsilon}}{3|g|} \right)^{\frac{1}{2}}
\int\limits_{-\infty}^{+\infty} d \vartheta \exp \left[-\frac{1}{2} \left(
\frac{16\pi^2 \lambda^{\varepsilon}}{3|g|} \, 2\delta \right) \vartheta^2
\right] =\frac{1}{\sqrt{2 \delta}}
\end{displaymath}
showing the spurious would-be zero-mode singularity, as the consequence of
false rotation-mode treatment.

Excluding the $l=0$~-mode in the numerator on the right-hand
side of (\ref{Tprod}),
we obtain for $c_2^T$:
\begin{eqnarray}
& \displaystyle{c_2^T}= & \!\!\! 2^{1/2} \exp \left\{-\frac{1}{2}
                          \sum_{l=1}^{\infty}
                          \frac{(2l+d-1) \Gamma(l+d-1)}{\Gamma(d)
                          \Gamma(l+1)} \ln \left[
                          \frac{(l+\frac{d}{2}+1)(l+\frac{d}{2}-2)}
                          {(l+\frac{d}{2}-1)(l+\frac{d}{2})}
                          \right] \nonumber \right. \\
                          \nonumber \\
&                       & \left.
                          -\frac{1}{2} \sum_{l=1}^{\infty}
                          \frac{(2l+d-1) \Gamma(l+d-1)}{
                          \Gamma(d) \Gamma(l+1)} \ln \left[
                          1+\frac{2\, \delta}{(l+\frac{d}{2}+1)
                          (l+\frac{d}{2}-2)} \right] \right\}.
                          \nonumber \\
\end{eqnarray}
The first sum is the well-known contribution from the isotropic limit. The
second sum is to be expanded in powers of $\delta$. For large $l$, the sum
diverges as $\varepsilon$ tends to zero.
In order to separate off the divergence
as a simple $1/{\varepsilon}$-pole, we use the
zeta-function regularization method described in \cite{Wal}.
A somewhat tedious but straightforward calculation leads to
\begin{equation}
c_2^T=(2\pi)^{-1/6} 3^{1/2} \exp \Bigg[\frac{1}{3\varepsilon}(1-\delta)^2
+\frac{1}{4}-\frac{1}{2}\, \gamma +
\frac{\zeta^{'} (2)}{\pi^2}-\frac{1}{9} \delta+
\frac{37}{81} {\delta}^2 +{\cal O} \left({\delta}^3 \right) \Bigg]\, .
\end{equation}
The series in front of the $\varepsilon$-singularity ends after the
quadratic power of $\delta$, thereby leading to the exact
$(1-\delta)^2/3$-coefficient of the simple $\varepsilon$-pole.

Collecting all contributions to the one-loop
expression of ${\rm Im} \, \tilde{G}^{(2M)}$,
we obtain the imaginary parts
\begin{eqnarray}
\lefteqn{{\rm Im} \, \tilde{G}^{(2M)} (x_1,x_2,
\ldots,x_{2M})_{i_1,i_2,\ldots,i_{2M}} = }
\nonumber \\
\nonumber \\ \
    & & \!\!\!\!\!\!\!\!
       -\int d \lambda \, d^dx_0 \left\{ c_1^L c_1^T c_2^L c_2^T
       (2\pi)^{-\frac{(d+1)}{2}} (2\pi)^{-1/2} J^{V_L} J^{V_T}
       \exp \left(\frac{\lambda^{\varepsilon}A}{g} \right)
       \nonumber \right. \\
       \nonumber \\
    & & \left. \mbox{} \!\!\!\!\!\!\!\!
       \times \prod^{2M}_{\nu=1} \phi_c(x_{\nu}) \,\,
       \frac{1}{2} \int\limits_0^{2\pi} d \vartheta u_{L\,i_1}(\vartheta)
       u_{L\,i_2}(\vartheta) \cdots u_{L\,i_{2M}} (\vartheta)
       \exp[a\sin^2(2\vartheta)] \right\}
\end{eqnarray}
with
\begin{equation}
\label{aA}
a=\frac{\lambda^{\varepsilon} 8\pi^2}{3g} \, \frac{\delta}{2} \, , \quad
A=\frac{8\pi^2}{3} \left[1-\frac{\varepsilon}{2}(2+\ln \pi+\gamma) \right]
\end{equation}
and $i_k=1,2$. The expression contains a standard factor $1/2$,
since for symmetry reasons each saddle point contributes
only one half of the Gaussian integral.

It remains to perform the collective coordinates
integration over the dilatation
$(\lambda)$, translation $(x_0)$, and the rotation
$(\vartheta)$ degree of freedom.
Having obtained the imaginary parts of the Green functions, we go over
to the bare imaginary part of the vertex functions.
Taking the Fourier transform and excluding the
$(2\pi)^d\, \delta(\sum_i q_i)$-factors, an amputation of the external legs
of $\tilde{G}^{(2M)}(x_1,x_2,\ldots,x_{2M})_{i_1,i_2,\ldots,i_{2M}}$
leads to:
\begin{eqnarray}
\label{V2M}
\lefteqn{{\rm Im} \,\Gamma_{\rm b}^{(2M)} (q_m)_{i_1,i_2,\ldots,i_{2M}} = }
\nonumber \\
\nonumber \\
    & &\!\!\!\!\!
       \displaystyle{-c_{\rm b}} \int\limits_0^{\infty}
       \frac{d\lambda}{\lambda} \left\{
       \lambda^{d-M(d-2)} \left(-\frac{\lambda^{\varepsilon}8\pi^2}{3g}
       \right)^{(d+2+2M)/2} \!\!\! \exp \left(\frac{\lambda^{\varepsilon}A}{g}
       \right) \prod^{2M}_{\nu=1}
       \left(\frac{q_{\nu}}{\lambda}\right)^2 \!
       \tilde{\phi} \left( \frac{q_{\nu}}{\lambda} \right)
       \nonumber \right. \\
       \nonumber \\
    & & \left. \mbox{} \!\!\!\!\!
       \times \frac{1}{2} \int\limits_0^{2\pi}
       d \vartheta u_{L\,i_1}(\vartheta)
       u_{L\,i_2}(\vartheta) \cdots u_{L\,i_{2M}} (\vartheta)
       \exp[a\sin^2(2\vartheta)] \right\} [1+{\cal O}(\varepsilon,g)]
\end{eqnarray}
with
\begin{equation}
\tilde{\phi}(q)=2^{d/2} \pi^{\frac{(d-2)}{2}} 3^{1/2} \left|q\right|^{1-(d/2)}
K_{\frac{d}{2}-1} (\left|q\right|) \, ,
\end{equation}
where $K_\mu(z)$ is the  modified Bessel-function.
The coefficient $c_{\rm b}$ is given by
\begin{eqnarray}
&c_{\rm b}=2^{-2/3} 3^{1/2} \pi^{-11/3} \exp &\!\!\!
                                \Bigg[
                                \frac{1}{3\varepsilon}
                                \left(10-2\delta+{\delta}^2\right)+
                                \frac{4 \zeta^{'} (2)}{\pi^2}-
                                4\gamma -4+\frac{8}{9} \, \delta
                                \nonumber  \\
                                \nonumber \\
&                               &\mbox{}
                                -\frac{7}{162} \, {\delta}^2
                                +{\cal O} \left( {\delta}^3 \right) \Bigg].
\end{eqnarray}
An $\varepsilon$-singularity results from the ultraviolet
divergence of one-loop diagrams. As discussed in the following section, this
simple pole in $\varepsilon$ is removed by a conventional coupling constant
renormalization.

The contribution from the imaginary part of the denominator in
equation (\ref{G2M}) follows for $M=0$ in (\ref{V2M}). We observe that it is
of higher order in $g$ in comparison with the imaginary part of the numerator,
and can therefore be neglected. So the imaginary part of (\ref{TilG})
supplies the desired imaginary part of (\ref{G2M}).
\section{Renormalization}
In the preceding section we have calculated the contribution of the quadratic
fluctuations around the isotropic instanton. The almost-zero eigenvalues of
translation, dilatation and rotation have been extracted and treated
separately. Of course, the resulting expressions
are useless, if we are not able
to renormalize the theory.
This is the most difficult additional problem which
arises in the transition from quantum mechanics to higher-dimensional field
theories.

A systematic scheme to renormalize both the real and imaginary part
of vertex functions for a $g{\phi}^4$ theory in $4-\varepsilon$ dimensions
was introduced by McKane, Wallace and de Alcantara Bonfim.
They calculated the full (real and imaginary)
renormalization-group constants using an extended minimal
subtraction scheme to one loop (the conventional
one is given in \cite{HoVe,Hooft}).
We have extended their method to
the case of two coupling constants in view of applications to critical
phenomena.

For an illustration we consider first the four-point vertex function.
The result (\ref{V2M}) of the one-loop calculation about the instanton reads:
\begin{eqnarray}
\label{pol}
\lefteqn{{\rm Im} \, \Gamma_{\rm b}^{(4)}(q_m)_{i,j,k,l}= } \nonumber \\
              \nonumber \\
&             &\displaystyle{\sum_{n=0}^{\infty}}
              \left\{ \frac{(-{\delta}/2)^n}{n!}
              \frac{1}{2} \int\limits_0^{2\pi} d \vartheta u_{L\,i}(\vartheta)
              u_{L\,j}(\vartheta) u_{L\,k}(\vartheta) u_{L\,l}(\vartheta)
              [\sin(2\vartheta)]^{2n}
              \nonumber \right. \\
              \nonumber \\
&             &\left. \mbox{} \times \left[
              -c_{\rm b} \int\limits_0^{\infty} \frac{d\lambda}{\lambda}
              \lambda^{\varepsilon}
              \left(-\frac{\lambda^{\varepsilon}A}{g}\right)^{\frac{d+6+2n}{2}}
              \!\!\! \exp \left(\frac{\lambda^{\varepsilon}A}{g}
              \right) \prod^{4}_{\nu=1}
              \left(\frac{q_{\nu}}{\lambda}\right)^2 \!
              \tilde{\phi}
              \left( \frac{q_{\nu}}{\lambda} \right) \right] \right\}
\end{eqnarray}
with $A=\frac{8{\pi}^2}{3} +{\cal O}(\varepsilon)$ and $i,j,k,l=1,2$.
It remains to evaluate the integral over the parameter $\lambda$. This is a
relict of the introduction of collective coordinates in the instanton
calculation. The integral converges for small $\lambda$ due to the
exponentially decreasing of modified Bessel-function $K_{d/2-1}$.
For large $\lambda$, the product
$(q_{\nu}/{\lambda})^2 \tilde{\phi}(q_{\nu}/{\lambda})$ behaves like
\begin{eqnarray}
\left(\frac{q_{\nu}}{\lambda}\right)^2 \!
\tilde{\phi}\left(\frac{q_{\nu}}{\lambda}\right)
\!\!&=&\!\!2^2 3^{1/2} \pi \Bigg\{
       \left(\frac{q_{\nu}}{2\lambda}\right)^{\varepsilon}
       \left[1+\frac{1}{2}\varepsilon \left(\gamma-\ln \pi \right)+
       {\cal O}\left({\varepsilon}^2\right) \right]
\nonumber \\
& &+\left(\frac{q_{\nu}}{2\lambda}\right)^2 \Bigg[
    \left(\frac{q_{\nu}}{2\lambda}\right)^{\varepsilon}
    \left(\frac{2}{\varepsilon}+\gamma-\ln \pi+{\cal O}(\varepsilon)\right)
\nonumber \\
& &\quad
    -\left(\frac{2}{\varepsilon}+1-\gamma-\ln \pi
    +{\cal O}(\varepsilon) \right) \Bigg]
    +{\cal O}\left(\frac{q_{\nu}}{2\lambda}\right)^4 \Bigg\} \, ,
\end{eqnarray}
so that the integral diverges logarithmically as $\varepsilon$ goes to zero.
This divergence causes a $1/{\varepsilon}$-pole in the imaginary part.
Since after the above approximation of
$\tilde{\phi}\left(q_{\nu}/{\lambda}\right)$ the integral diverges for
$\lambda \rightarrow 0$, a small $\lambda$ cutoff $\mu$ has to be
introduced.
In this way, the $\lambda$-integral in (\ref{pol}) takes the form
\begin{equation}
\label{scal}
-c_{\rm b} \int\limits_{\mu}^{\infty} \frac{d\lambda}{\lambda}
\lambda^{\varepsilon}
\left(-\frac{\lambda^{\varepsilon}A}{g}\right)^{\frac{d+6+2n}{2}}
\!\!\! \exp \left(\frac{\lambda^{\varepsilon}A}{g} \right)
(2^2 3^{1/2} \pi)^4 \prod^{4}_{\nu=1} \left(\frac{q_{\nu}}{2\lambda}
\right)^{\varepsilon}
\left[1+{\cal O}\left(\frac{q_{\nu}}{2\lambda} \right) \right] \, .
\end{equation}
The coupling constant $g$ is chosen to lie on top of the tip of the left-hand
cut in the complex $g$-plane. Being integrated only near
$g \rightarrow 0^{-}$, the integral can be evaluated perturbatively in $g$.
Using
\begin{equation}
\label{lambda}
\int\limits_{\mu}^{\infty} \frac{d\lambda}{\lambda} \lambda^{a\varepsilon}
\lambda^{\varepsilon}
\exp\left[-\left(\frac{\lambda^{\varepsilon} A}{|g|}\right)\right]=
\frac{1}{\varepsilon}\frac{|g|}{A} \mu^{a\varepsilon}
\exp\left[-\left(\frac{\mu^{\varepsilon} A}{|g|}\right)\right]
[1+{\cal O}(g)]\, ,
\end{equation}
we get for (\ref{scal}) the expression
\begin{equation}
2^8 3^2 \pi^4 \,c_{\rm b}\,\, \frac{1}{\varepsilon}\,\, \frac{g}{A}\,
\left(-\frac{\mu^{\varepsilon}A}{g}\right)^{\frac{d+6+2n}{2}}
\!\!\! \exp \left(\frac{\mu^{\varepsilon}A}{g} \right)
\prod^{4}_{\nu=1} \left(\frac{q_{\nu}}{\mu}
\right)^{\varepsilon}.
\end{equation}
After expanding the factor $(q_{\nu}/{\mu})^{\varepsilon}$ in powers of
$\varepsilon$, we obtain
the typical finite contribution $\ln(q_{\nu}/{\mu})$. Since this term can be
omitted by a minimal subtraction, we are left with
\begin{eqnarray}
\label{gim}
\lefteqn{{\rm Im} \, \Gamma_{\rm b}^{(4)}(q_m)_{i,j,k,l}= } \nonumber \\
              \nonumber \\
&             &2^8 3^2 \pi^4 \,c_{\rm b}\,\, \frac{1}{\varepsilon}
              \sum_{n=0}^{\infty}
              \left\{ \frac{(-2\, \delta)^n}{n!\Gamma(2n+3)}
              \left(\frac{g}{A}\right)
              \left(-\frac{\mu^{\varepsilon}A}{g}\right)^{\frac{d+6+2n}{2}}
              \!\!\! \exp \left(\frac{\mu^{\varepsilon}A}{g} \right)
              \nonumber \right. \\
              \nonumber \\
&             &\left. \mbox{} \times \left[
              3\Gamma\left(n+\frac{3}{2}\right)^2 \,\,(S_{ijkl}-\delta_{ijkl})+
              \Gamma\left(n+\frac{1}{2}\right) \Gamma\left(n+\frac{5}{2}\right)
              \,\, \delta_{ijkl}
              \right] \right\}\!,
\end{eqnarray}
where
\begin{displaymath}
S_{ijkl}=\frac{1}{3} (\delta_{ij} \delta_{kl}+\delta_{ik} \delta_{jl}+
\delta_{il} \delta_{jk})
\end{displaymath}
and
\begin{displaymath}
\delta_{ijkl} = \left\{ \begin{array}{l}
                1\, , \quad i=j=k=l \\ 0\, , \quad {\rm otherwise}\, .
                 \end{array} \right. \qquad (i,j,k,l=1,2)
\end{displaymath}
The tensor structure arises explicitly upon doing the integrals
\begin{displaymath}
\int\limits_0^{{\pi}/2}
\sin^{2\alpha+1} (\vartheta) \cos^{2\beta+1} (\vartheta)
\,d\vartheta = \frac{\Gamma(\alpha+1) \Gamma(\beta+1)}{2\,\,
\Gamma(\alpha+\beta+2)}=\frac{1}{2} B(\alpha+1,\beta+1)
\end{displaymath}
for all combinations of indices in ${\rm Im} \,{\Gamma}_{ijkl}^{(4)}$.
The first term in (\ref{gim}) contains only mixed index combinations,
whereas the second term has only a contribution from equal indices.

Now we can proceed to
renormalize the bare results.
The real part of ${\Gamma}_{ijkl}^{(4)}(q)$ is the
perturbative one, and it is easily calculated to one loop. In terms of our
coupling constants,
the real parts ${\rm Re} \,{\Gamma}_{ijkl}^{(4)}(q)$ take the form
\begin{eqnarray}
\label{greal}
\lefteqn{{\rm Re} \, \Gamma_{\rm b}^{(4)}(q_m)_{i,j,k,l}= } \nonumber \\
              \nonumber \\
&             &\!\!\!\!\!
              \left\{-6(1-\delta) g+\left[\frac{3}{4\pi^2\varepsilon}
              \left(10-14 \delta+4 {\delta}^2\right)
              +{\cal O}({\varepsilon}^0) \right]
              g^2 \mu^{-\varepsilon} \right\} \{
              S_{ijkl}-\delta_{ijkl} \}
              \nonumber \\
              \nonumber \\
&             &\!\!\!\!\!
              +\left\{-6g+\left[\frac{3}{4\pi^2\varepsilon}
              \left(10-2 \delta+{\delta}^2\right)
              +{\cal O}({\varepsilon}^0) \right]
              g^2 \mu^{-\varepsilon} \right\}
              \, \delta_{ijkl}\, ,
\end{eqnarray}
where $\mu$ is the arbitrary momentum scale introduced above in (\ref{scal}).

In the absence of an instanton, the wave function renormalization has no
one-loop contribution. In the consequence,
the expression (\ref{greal})
is rendered finite by a coupling constant renormalization
only. The subscript r of the coupling constants indicates
renormalized expressions in the conventional perturbative manner, i.~e.,
\begin{eqnarray}
g_{\rm r}\!\!\!&=&\!\!\!g\mu^{-\varepsilon}-\frac{1}{8\pi^2\varepsilon}
     \left(10-2\delta+{\delta}^2 \right) g^2 \mu^{-2\varepsilon}\, ,
     \nonumber\\
     \nonumber\\
\delta_{\rm r}\!\!&=&\!\!\delta+\frac{1}{8\pi^2\varepsilon}
        \left(-2\delta+{\delta}^2+{\delta}^3 \right) g \mu^{-\varepsilon}
         \nonumber \\
\end{eqnarray}
Inserting this into (\ref{gim}) and (\ref{greal}) we find
the perturbatively renormalized vertex functions
\begin{equation}
\label{gren}
\Gamma_{\rm r}^{(4)}(q_m)_{i,j,k,l}=-6\mu^{\varepsilon}
\left[F_1(\delta_{\rm r},g_{\rm r})
\left(S_{ijkl}-\delta_{ijkl}\right)+F_2(\delta_{\rm r},g_{\rm r})\delta_{ijkl}
\right]
\end{equation}
with
\begin{eqnarray}
\lefteqn{F_1({\delta}_{\rm r},g_{\rm r})= } \nonumber \\
 & &  \!\!\!\!\!(1-{\delta}_{\rm r}) g_{\rm r}+i 2^7 3^2 \pi^4 \,
      c_{\rm r} \,\, \frac{1}{\varepsilon}
      \sum_{n=0}^{\infty} \frac{(-2{\delta}_{\rm r})^n}{n!}
      \frac{\Gamma\left(n+\frac{3}{2} \right)^2}{\Gamma(2n+3)}
      \left(-\frac{A}{g_{\rm r}}\right)^{\frac{d+4+2n}{2}}
      \!\!\! \exp \left(\frac{A}{g_{\rm r}} \right),
      \nonumber \\
\end{eqnarray}
and
\begin{eqnarray}
&F_2&\!\!\!\!({\delta}_{\rm r},g_{\rm r})=  \nonumber \\
&g_{\rm r}&\!\!\!\!+i 2^7 3 \pi^4 \,
     c_{\rm r} \, \frac{1}{\varepsilon}
     \sum_{n=0}^{\infty} \frac{(-2{\delta}_{\rm r})^n}{n!}
     \frac{\Gamma\left(n+\frac{1}{2} \right) \Gamma\left(n+\frac{5}{2} \right)}
     {\Gamma(2n+3)}
     \left(-\frac{A}{g_{\rm r}}\right)^{\frac{d+4+2n}{2}}
     \!\!\! \exp \left(\frac{A}{g_{\rm r}} \right).
     \nonumber \\
\end{eqnarray}
The coefficient $c_{\rm r}$ is given by
\begin{eqnarray}
\label{cr}
c_{\rm r}\!\!\! &=&\!\!\! c_{\rm b}
        \exp \left[ -\frac{A}{8 \pi^2\varepsilon}
        \left(10-2 {\delta}_{\rm r}
        +{\delta}_{\rm r}^2 \right) \right]
        \nonumber \\
        \nonumber \\
    &=&\!\!\!  3^{1/2} 2^{-2/3} \pi^{-2} \exp \bigg[\frac{1}{3\varepsilon}
        \left(10-2 {\delta}_{\rm r}+{\delta}_{\rm r}^2 \right)-
        \frac{1}{3\varepsilon}
        \left(10-2 {\delta}_{\rm r}+{\delta}_{\rm r}^2 \right)
        \nonumber \\
        \nonumber \\
    & & \qquad \qquad \qquad \quad
        +\frac{4\zeta^{'}(2)}{\pi^2}-\frac{7}{3}\gamma-\frac{2}{3}
        +{\cal O}({\delta}_{\rm r})
        +{\cal O}({\delta}_{\rm r} \, g_{\rm r}) \bigg].
\end{eqnarray}
The $1/{\varepsilon}$-pole terms cancel.
Thus, $c_{\rm r}$ remains finite for $\varepsilon \rightarrow 0$.
For the leading expansion (\ref{IMG}) we can
take the perturbative renormalized
expression (\ref{cr}) at the position ${\delta}_{\rm r} = 0$. The only
remaining singularity
is the $1/{\varepsilon}$-factor in
${\rm Im} \,{\Gamma}^{(4)}_{\rm r}(q)_{ijkl}$.
It requires a further renormalization.
In our special choice of coupling constants, the disastrous
$k!$-divergence of perturbation series appears in
the expansions coefficients of $g^k$.
Since $\delta$ is a
dimensionless anisotropy measure for positive as well as negative $g$,
we have expanded the imaginary part of vertex functions around the isotropic
case in a well defined manner. For the derivation of nonperturbative
corrections to the renormalization-group functions and for later
applications, however, it is convenient
to avoid ratios of coupling constants. Therefore we proceed
with $v_{\rm r} = g_{\rm r} \, {\delta}_{\rm r}$ and
$g_{\rm r}$ instead of ${\delta}_{\rm r}$ and $g_{\rm r}$.
Then $v_{\rm r}$ receive the rule of ${\delta}_{\rm r}$.

Now we apply the extended minimal subtraction and perform a second
(nonperturbative) renormalization step to eliminate
the $1/{\varepsilon}$-pole term:
\begin{eqnarray}
\label{REN}
g_{\rm R}\!\!&=&\!\!g_{\rm r}+i 2^7 3 \pi^4 c_{\rm r} \frac{1}{\varepsilon}
\displaystyle{\sum_{n=0}^{\infty}}
\Bigg[ \frac{v_{\rm r}^n}{n!} \left(\frac{3}{4\pi^2}\right)^n
B \left( n+\frac{1}{2},n+\frac{5}{2} \right)
\nonumber  \\
& & \qquad \qquad \qquad \quad
\times \left(-\frac{A}{g_{\rm r}} \right)^{\frac{d+4+4 n}{2}}
\exp \left(\frac{A}{g_{\rm r}} \right) \Bigg]\, , \nonumber \\
\nonumber \\
\nonumber \\
v_{\rm R}\!\!&=&\!\!v_{\rm r}-i 2^7 3 \pi^4 c_{\rm r} \frac{1}{\varepsilon}
\displaystyle{\sum_{n=0}^{\infty}}
\Bigg[ \frac{v_{\rm r}^n}{n!} \left(\frac{3}{4\pi^2}\right)^n
\left(\frac{4n}{2n+3}\right)
B \left(n+\frac{1}{2},n+\frac{5}{2} \right)
\nonumber  \\
& & \qquad \qquad \qquad \quad
\times \left(-\frac{A}{g_{\rm r}} \right)^{\frac{d+4+4 n}{2}}
\exp \left(\frac{A}{g_{\rm r}} \right) \Bigg]\, .
\end{eqnarray}
The subscript R denotes the fully renormalized coupling constants. The
required wave function renormalization in nonperturbative terms give no
leading contributions to the imaginary part (see below). Therefore the
$1/{\varepsilon}$-singularity in (\ref{gren}) is removed by a
nonperturbative coupling-constant renormalization only.

We now proceed to investigate the two-step renormalization for the
wave function. There are two additional complications for the
${\Gamma}^{(2)}$-functions. First, ${\rm Im} \,{\Gamma}^{(2)}$ is quadratically
divergent. Second, there is a momentum-dependent divergence in the imaginary
part of ${\Gamma}^{(2)}$. For the $g{\phi}^4$-theory it was shown in
\cite{WalBon} that the undesirable momentum-dependence disappears
during the nonperturbative renormalization process. This is comparable to the
situation in conventional two-loop perturbation expansion. We continue to
show that this non-obvious cancellation still works in the more complex case
of two coupling constants and more than one field components.

According to (\ref{V2M}) the imaginary part of
the unrenormalized two-point vertex function reads
\begin{eqnarray}
{\rm Im} \, \Gamma_{\rm b}^{(2)}(q)_{i,j}\!\!&=&\!\!
              c_{\rm b} \int\limits_0^{\infty} \frac{d\lambda}{\lambda}
              \left\{ \lambda^2
              \left(-\frac{\lambda^{\varepsilon}A}{g}\right)^{\frac{d+4}{2}}
              \!\!\! \exp \left(\frac{\lambda^{\varepsilon}A}{g} \right)
              \left[ \left(\frac{q}{\lambda}\right)^2 \!
              \tilde{\phi} \left( \frac{q}{\lambda} \right) \right]^2
              \nonumber \right.\\
              \nonumber \\
&             &\left. \qquad \qquad \times
              \frac{1}{2} \int\limits_0^{2\pi} d \vartheta u_{L\,i}(\vartheta)
              u_{L\,j}(\vartheta)
              \exp\left[a \sin^2(2\vartheta)\right] \right\}.
\end{eqnarray}
The integration over the angle $\vartheta$ can be done:
\begin{eqnarray}
\lefteqn{\frac{1}{2} \int\limits_0^{2\pi}
d\vartheta u_{L\,i}(\vartheta)
u_{L\,j}(\vartheta) \exp\left[a \sin^2(2\vartheta)\right]}
\nonumber \\
\nonumber \\
& &\!\!\!\!\!\!\!
=\delta_{ij} \sum_{n=0}^{\infty} \frac{a^n}{n!}\, 4^n \frac{\Gamma\left(n+
\frac{1}{2} \right) \Gamma\left(n+\frac{3}{2} \right) }{\Gamma(2n+2)}
=\delta_{ij} \sum_{n=0}^{\infty} \frac{a^n}{n!}\, 4^n B \left(n+
\frac{1}{2},n+\frac{3}{2} \right).
\nonumber \\
\end{eqnarray}
Thus, we get the usual form:
${\rm Im} \,{\Gamma}^{(2)}_{b\,\,ij}
={\delta}_{ij}\,{\rm Im} \,{\Gamma}^{(2)}_{\rm b}$.
The wave function renormalization is contained in
$\frac{\partial}{\partial q^2} {\Gamma}^{(2)}(q)$.
Following a similar procedure
as in the case of ${\Gamma}^{(4)}$, we obtain the imaginary part
\begin{eqnarray}
\label{WFim}
\frac{\partial}{\partial q^2}\, {\rm Im} \,\Gamma_{\rm b}^{(2)}(q)\!\!&=&\!\!
2^4 \pi^2 3 \,c_{\rm b}\,
\frac{1}{\varepsilon} \sum_{n=0}^{\infty} \left\{\frac{(-2\, \delta)^n}{n!}
B \left(n+\frac{1}{2},n+\frac{3}{2} \right)
\nonumber \right. \\
\nonumber \\
& &\left. \mbox{} \times
\left[\gamma-\frac{1}{2}+\ln \left(\frac{q}{2\mu}\right) \right]
\left(-\frac{\mu^{\varepsilon}A}{g} \right)^{\frac{d+2+2n}{2}} \exp\left(
\frac{\mu^{\varepsilon}A}{g}\right) \right\} \, .
\nonumber \\
\end{eqnarray}
Apart from the appearance of the $(1/{\varepsilon}) \ln(q/{\mu})$ term,
a very similar expression is found for
${\rm Im} \,{\Gamma}^{(4)}_{\rm b}(q)$. In order to
show the cancellation of this term during the second stage of renormalization
we need the two-loop expression for the real part of
$\frac{\partial}{\partial q^2} {\Gamma}^{(2)}(q)$
\begin{equation}
\label{WFreal}
\frac{\partial}{\partial q^2}\,{\rm Re}\,\Gamma_{\rm b}^{(2)}(q)=
1+q^{-2\varepsilon}
\left(\frac{4}{3}-\frac{2}{3} \delta+\frac{1}{3} {\delta}^2 \right) g^2
\left[\frac{3}{4(8\pi^2)^2\varepsilon}
+{\rm P}+{\cal O}(\varepsilon) \right] \, ,
\end{equation}
where {\rm P} denotes some number which is the contribution of
order ${\varepsilon}^0$.
At $q={\mu}$, (\ref{WFreal}) is commonly defined as the inverse wave
function renormalization constant $(Z^{\phi}_{\rm p})^{-1}$. In terms of the
perturbative renormalized coupling constants $Z^{\phi}_{\rm p}$ is given by
\begin{equation}
\label{zp}
Z_{\rm p}^{\phi}=1-\frac{1}{4(8\pi^2)^2\varepsilon}
\left(4-2 {\delta}_{\rm r}
+{\delta}_{\rm r}^2 \right) g_{\rm r}^2  \, ,
\end{equation}
and together with (\ref{WFim}), we find after the first step of renormalization
\begin{eqnarray}
\frac{\partial}{\partial q^2}\,\Gamma_{\rm r}^{(2)}(q)\!\!&=&\!\!
1+\left(\frac{4}{3} g_{\rm r}^2-\frac{2}{3} v_{\rm r} g_{\rm r}
+\frac{1}{3} v_{\rm r}^2 \right)
\left[{\rm P}-\frac{3}{2(8\pi^2)^2} \ln \left(\frac{q}{\mu}\right)
+{\cal O}(\varepsilon)
\right]
\nonumber \\
\nonumber \\
&+&\!\!i\,2^4 \pi^2 3 \,c_{\rm r}\,
\frac{1}{\varepsilon} \sum_{n=0}^{\infty} \left\{\frac{v_{\rm r}^n}{n!}
\left(\frac{3}{4\pi^2}\right)^n
B \left(n+\frac{1}{2},n+\frac{3}{2} \right)
\nonumber \right. \\
\nonumber \\
& &\left. \mbox{} \times
\left[\gamma-\frac{1}{2}+\ln \left(\frac{q}{2\mu}\right) \right]
\left(-\frac{A}{g_{\rm r}} \right)^{\frac{d+2+4n}{2}} \exp\left(
\frac{A}{g_{\rm r}}\right) \right\} \, .
\end{eqnarray}
We have again introduced the coupling constants $v_{\rm r}$ and $g_{\rm r}$
which are more convenient for practical applications.
This expression reads in the fully renormalized
couplings (\ref{REN})
\begin{eqnarray}
\lefteqn{\frac{\partial}{\partial q^2}\,\Gamma_{\rm r}^{(2)}(q) =
1+\left(\frac{4}{3} g_{\rm R}^2-\frac{2}{3} v_{\rm R} g_{\rm R} +
\frac{1}{3} v_{\rm R}^2 \right)
\left({\rm P}-\frac{3}{2(8\pi^2)^2} \ln \left(\frac{q}{\mu}\right)
\right)} \nonumber \\
\nonumber \\
& & \mbox{}+i\,2^4 \pi^2 3 \,c_{\rm r}\,
\frac{1}{\varepsilon} \sum_{n=0}^{\infty} \Bigg\{\frac{v_{\rm R}^n}{n!}
\left(\frac{3}{4\pi^2}\right)^n
\nonumber \\
\nonumber \\
& & \qquad
\times \Bigg[2^4 \pi^2 A \,\,
B \left(n+\frac{1}{2},n+\frac{5}{2} \right)\, \frac{4(n+1)}{2n+3}
\left({\rm P}-\frac{3}{2(8\pi^2)^2} \ln \left(\frac{q}{\mu}\right) \right)
\nonumber \\
\nonumber \\
& & \qquad \quad
+B \left(n+\frac{1}{2},n+\frac{3}{2} \right)
\left(\gamma-\frac{1}{2}-\ln2+\ln \left(\frac{q}{\mu}\right) \right) \Bigg]
\nonumber \\
\nonumber \\
& &
\times \left(-\frac{A}{g_{\rm R}} \right)^{\frac{d+2+4n}{2}} \exp\left(
\frac{A}{g_{\rm R}}\right) \Bigg\} \, .
\end{eqnarray}
By using the definition of the $B$-function in terms of Gamma-functions one
can easily read off the cancellation of $(1/{\varepsilon}) \ln(q/{\mu})$
singularities for every power $n$ of the coupling constant $v_{\rm R}$.
Therefore we
are left with
\begin{eqnarray}
\frac{\partial}{\partial q^2}\,\Gamma_{\rm r}^{(2)}(q)\!\!&=&\!\!
1+\left(\frac{4}{3} g_{\rm R}^2-\frac{2}{3} v_{\rm R} g_{\rm R}
+\frac{1}{3} v_{\rm R}^2 \right)
\left[{\rm P}-\frac{3}{2(8\pi^2)^2} \ln \left(\frac{q}{\mu}\right)
\right]
\nonumber \\
\nonumber \\
&+&\!\! i\,2^4 \pi^2 3 \,c_{\rm r}\,
\frac{1}{\varepsilon} \left(2^4 \pi^2 A\,\,{\rm P}+\gamma-\frac{1}{2}-\ln2
\right)
\nonumber \\
\nonumber \\
&\displaystyle{\times}&\!\!\!
\sum_{n=0}^{\infty} \Bigg[\frac{v_{\rm R}^n}{n!}
\left(\frac{3}{4\pi^2}\right)^n B \left(n+\frac{1}{2},n+\frac{3}{2} \right)
\left(-\frac{A}{g_{\rm R}} \right)^{\frac{d+2+4n}{2}} \exp\left(
\frac{A}{g_{\rm R}}\right) \Bigg] \, .
\nonumber \\
\end{eqnarray}
Now we can define the nonperturbative minimally subtracted wave function
renormalization to remove the $1/{\varepsilon}$-pole:
\begin{eqnarray}
\label{znp}
Z_{{\rm np}}^{\phi}\!\!&=&\!\!1-
i\,2^4 \pi^2 3 \,c_{\rm r}\,
\frac{1}{\varepsilon} \left(2^4 \pi^2 A\,\,{\rm P}+\gamma-\frac{1}{2}-\ln2
\right)
\nonumber \\
\nonumber \\
& &\qquad \times \sum_{n=0}^{\infty} \Bigg[\frac{v_{\rm R}^n}{n!}
\left(\frac{3}{4\pi^2}\right)^n B \left(n+\frac{1}{2},n+\frac{3}{2} \right)
\nonumber \\
\nonumber \\
& &\qquad \qquad \quad
\times \left(-\frac{A}{g_{\rm R}} \right)^{\frac{d+2+4n}{2}} \exp\left(
\frac{A}{g_{\rm R}}\right) \Bigg] \, .
\end{eqnarray}
The full renormalization constant is defined by
$Z^{\phi}:=Z^{\phi}_{\rm p}\,Z^{\phi}_{{\rm np}}$ and follows
from the product of (\ref{zp}) and (\ref{znp}).

A further important Green function is associated with the composite
field $(1/2) {\phi}^2$:
\begin{equation}
G^{(2,1)}(x_1,x_2;x_3)_{i,j}=\left\langle {\phi}_i(x_1) {\phi}_j(x_2)
\frac{1}{2} {\phi}_k^2(x_3) \right\rangle \, .
\end{equation}
By a straightforward application of the techniques discussed in the
preceding sections we obtain for the corresponding vertex function
\begin{eqnarray}
\lefteqn{ {\rm Im} \, \Gamma_{\rm b}^{(2,1)}(q_1,q_2;q_3)_{i,j}=
         -\frac{1}{2} c_{\rm b} \int\limits_0^{\infty} \frac{d\lambda}{\lambda}
          \Bigg\{
          \left(-\frac{\lambda^{\varepsilon}A}{g}\right)^{\frac{d+6}{2}}
          \!\!\! \exp \left(\frac{\lambda^{\varepsilon}A}{g} \right)}
              \nonumber \\
              \nonumber \\
&             &\!\!\!\!\! \times
              \prod_{\nu=1}^2 \left[ \left(\frac{q_\nu}{\lambda}\right)^2 \!
              \tilde{\phi} \left( \frac{q_\nu}{\lambda} \right) \right]
              \hat{\phi} \left(\frac{q_3}{\lambda}\right)
              \frac{1}{2} \int\limits_0^{2\pi} d \vartheta u_{L\,i}(\vartheta)
              u_{L\,j}(\vartheta)
              \exp\left[a \sin^2(2\vartheta)\right] \Bigg\} \, ,
\nonumber \\
\end{eqnarray}
where
\begin{equation}
\hat{\phi}(q)=2^{(d-2)/2}\pi^{(d-4)/2} 3
\left| q \right|^{(4-d)/2} K_{(d-4)/2}\left(\left|q \right|\right) \, .
\end{equation}
Now we can proceed as in the case of wave function renormalization.
Performing the collective coordinates integration over $\lambda$ and
$\vartheta$ we obtain
\begin{eqnarray}
{\rm Im} \,\Gamma_{\rm b}^{(2,1)}(q_1,q_2;q_3)_{i,j}\!\!&=&\!\!
{\delta}_{ij} \, 2^4 \pi^2 3^2 \,c_{\rm b}\,
\frac{1}{\varepsilon} \sum_{n=0}^{\infty} \left\{\frac{(-2\, \delta)^n}{n!}
B \left(n+\frac{1}{2},n+\frac{3}{2} \right)
\nonumber \right. \\
\nonumber \\
& &\left. \!\!\! \times
\left[\gamma+\ln \left(\frac{q_3}{2\mu}\right)\right]
\left(-\frac{\mu^{\varepsilon}A}{g} \right)^{\frac{d+4+2n}{2}} \exp\left(
\frac{\mu^{\varepsilon}A}{g}\right) \right\},
\nonumber \\
\end{eqnarray}
which agrees with the usual form ${\Gamma}^{(2,1)}_{{\rm b}\, ij}=
\delta_{ij}\, {\Gamma}^{(2,1)}_{{\rm b}}$.
At $q_3=0$, $q_1^2=q_2^2={\mu}^2$, the vertex function
${\Gamma}^{(2,1)}(q_1,q_2;q_3)$ defines a third renormalization constant,
called $Z^{\phi^2}$, via
\begin{equation}
{\Gamma}^{(2,1)} \equiv \left(Z^{\phi} Z^{\phi^2} \right)^{-1} \, .
\end{equation}
The perturbative $Z^{\phi^2}$ follows from the real part of
${\Gamma}^{(2,1)}$:
\begin{equation}
{\rm Re} \Gamma^{(2,1)}_{\rm b}=1+
g q_3^{-\varepsilon} \left(\frac{4}{3}-\frac{1}{3} \delta \right)
\left(-\frac{3}{8\pi^2\varepsilon}+{\rm Q}+{\cal O}(\varepsilon) \right)\, ,
\end{equation}
where ${\rm Q}$ is a number of order ${\varepsilon}^0$.
In terms of the perturbative renormalized coupling constants, the one-loop
expression of $Z_p^{\phi^2}$ is given by
\begin{equation}
\label{pPP}
Z_{\rm p}^{\phi^2}=1+\frac{1}{8\pi^2\varepsilon}
\left(4-\delta_{\rm r}\right) g_{\rm r} \, .
\end{equation}
Including ${\rm Re} \Gamma^{(2,1)}_{\rm b}$
and ${\rm Im} \Gamma^{(2,1)}_{\rm b}$
we obtain after the perturbative step of renormalization
\begin{eqnarray}
\Gamma_{\rm r}^{(2,1)}(q_1,q_2;q_3)\!\!&=&\!\!
1+\left(\frac{4}{3} g_{\rm r}-\frac{1}{3} v_{\rm r}\right)
\left[{\rm Q}+\frac{3}{8\pi^2} \ln \left(\frac{q_3}{\mu}\right)
+{\cal O}(\varepsilon)
\right]
\nonumber \\
\nonumber \\
&+&\!\!i\,2^4 \pi^2 3^2 \,c_{\rm r}\,
\frac{1}{\varepsilon} \sum_{n=0}^{\infty} \left\{\frac{v_{\rm r}^n}{n!}
\left(\frac{3}{4\pi^2}\right)^n
B \left(n+\frac{1}{2},n+\frac{3}{2} \right)
\nonumber \right. \\
\nonumber \\
& &\left. \mbox{} \times
\left[\gamma+\ln \left(\frac{q_3}{2\mu}\right) \right]
\left(-\frac{A}{g_{\rm r}} \right)^{\frac{d+4+4n}{2}} \exp\left(
\frac{A}{g_{\rm r}}\right) \right\} \, .
\end{eqnarray}
In the second step of renormalization the $(1/{\varepsilon})\ln (q/{\mu})$
singularities cancel, and we find
\begin{eqnarray}
&\Gamma&\!\!\!\!\!_{\rm r}^{(2,1)}(q_1,q_2;q_3)=
1+\left(\frac{4}{3} g_{\rm R}-\frac{1}{3} v_{\rm R} \right)
\left[{\rm Q}+\frac{3}{8\pi^2} \ln \left(\frac{q_3}{\mu}\right)
\right]
\nonumber \\
\nonumber \\
& &\quad -i\,2^4 \pi^2 3^2 \,c_{\rm r}\,
\frac{1}{\varepsilon} \left(\frac{8\pi^2}{3} {\rm Q}-\gamma+\ln2 \right)
\nonumber \\
\nonumber \\
& &\quad \displaystyle{\times}
\sum_{n=0}^{\infty} \Bigg[\frac{v_{\rm R}^n}{n!}
\left(\frac{3}{4\pi^2}\right)^n B \left(n+\frac{1}{2},n+\frac{3}{2} \right)
\left(-\frac{A}{g_{\rm R}} \right)^{\frac{d+4+4n}{2}} \exp\left(
\frac{A}{g_{\rm R}}\right) \Bigg] \, .
\nonumber \\
\end{eqnarray}
Hence, the nonperturbative $Z^{\phi^2}$ is
\begin{eqnarray}
\label{npPP}
\lefteqn{Z_{{\rm np}}^{\phi} Z_{{\rm np}}^{\phi^2}=1+
i\,2^4 \pi^2 3^2 \,c_{\rm r}\,
\frac{1}{\varepsilon} \left(\frac{8\pi^2}{3} {\rm Q}-\gamma+\ln2 \right)}
\nonumber \\
\nonumber \\
& &\times \sum_{n=0}^{\infty} \Bigg[\frac{v_{\rm R}^n}{n!}
\left(\frac{3}{4\pi^2}\right)^n B \left(n+\frac{1}{2},n+\frac{3}{2} \right)
\left(-\frac{A}{g_{\rm R}} \right)^{\frac{d+4+4n}{2}} \exp\left(
\frac{A}{g_{\rm R}}\right) \Bigg].
\end{eqnarray}
With respect to the leading expansion (\ref{IMG}) this is equal to
$Z_{{\rm np}}^{\phi^2}$.
\section{Renormalization group functions}
After having calculated the imaginary parts of the renormalized couplings
and of the renormalization constants $Z^{\phi}$ and $Z^{\phi^2}$,
we are in the position to derive the nonperturbative renormalization
group functions $\beta(g_{\rm R},v_{\rm R})$, $\gamma(g_{\rm R},v_{\rm R})$
and $\gamma_{\phi^2}(g_{\rm R},v_{\rm R})$ in the usual way. First we derive
the $\beta$-functions from the expressions (\ref{REN}) for $g_{\rm R}(\mu)$
and $v_{\rm R}(\mu)$, respectively, and obtain:
\begin{eqnarray}
\lefteqn{\beta_g (g_{\rm R},v_{\rm R})=} \nonumber \\
& &\mu \frac{\partial}{\partial \mu} g_{\rm R}
=\mu \frac{\partial}{\partial \mu} g_{\rm r}+
i 2^7 3 \pi^4 c_{\rm r} \frac{1}{\varepsilon}
\displaystyle{\sum_{n=0}^{\infty}} \mu \frac{\partial}{\partial \mu}
\Bigg[ \frac{v_{\rm r}^n}{n!} \left(\frac{3}{4\pi^2}\right)^n
B \left( n+\frac{1}{2},n+\frac{5}{2} \right)
\nonumber  \\
\nonumber \\
& & \qquad \qquad \qquad \qquad \qquad
\times \left(-\frac{A}{g_{\rm r}} \right)^{\frac{d+4+4 n}{2}}
\exp \left(\frac{A}{g_{\rm r}} \right) \Bigg]\, ,
\nonumber \\
\lefteqn{\beta_v(g_{\rm R},v_{\rm R})=} \nonumber \\
& &\mu \frac{\partial}{\partial \mu} v_{\rm R}
=\mu \frac{\partial}{\partial \mu} v_{\rm r}-
i 2^7 3 \pi^4 c_{\rm r} \frac{1}{\varepsilon}
\displaystyle{\sum_{n=0}^{\infty}} \mu \frac{\partial}{\partial \mu}
\Bigg[ \frac{v_{\rm r}^n}{n!} \left(\frac{3}{4\pi^2}\right)^n
B \left(n+\frac{1}{2},n+\frac{5}{2} \right)
\nonumber  \\
\nonumber \\
& & \qquad \qquad \qquad \qquad \qquad
\times \left(\frac{4n}{2n+3}\right)
\left(-\frac{A}{g_{\rm r}} \right)^{\frac{d+4+4 n}{2}}
\exp \left(\frac{A}{g_{\rm r}} \right) \Bigg]\, .
\end{eqnarray}
The derivatives are taken at fixed bare couplings $g$ and $v$. These
functions will be finite as $\varepsilon \rightarrow 0$. Furthermore, they
will have a left-hand cut in the complex $g$-plane. On top of the tip of
this cut we find
\begin{eqnarray}
\label{bet1}
\beta_g (g_{\rm R},v_{\rm R})=
\beta_g^{\rm p}(g_{\rm r},v_{\rm r})
\!\!\!\!&-&\!\!\!\!i 2^7 3 \pi^4 c_{\rm r}
\displaystyle{\sum_{n=0}^{\infty}}
\Bigg[ \frac{v_{\rm r}^n}{n!} \left(\frac{3}{4\pi^2}\right)^n
B \left( n+\frac{1}{2},n+\frac{5}{2} \right)
\nonumber \\
\nonumber \\
& &
\times \left(-\frac{A}{g_{\rm r}} \right)^{\frac{d+6+4 n}{2}}
\exp \left(\frac{A}{g_{\rm r}} \right) \Bigg] \, ,
\nonumber \\
\nonumber \\
\beta_v(g_{\rm R},v_{\rm R})=
\beta_v^{\rm p}(g_{\rm r},v_{\rm r})
\!\!\!\!&+&\!\!\!\!i 2^7 3 \pi^4 c_{\rm r}
\displaystyle{\sum_{n=0}^{\infty}}
\Bigg[ \frac{v_{\rm r}^n}{n!} \left(\frac{3}{4\pi^2}\right)^n
B \left(n+\frac{1}{2},n+\frac{5}{2} \right)
\nonumber  \\
\nonumber \\
& &
\times \left(\frac{4n}{2n+3}\right)
\left(-\frac{A}{g_{\rm r}} \right)^{\frac{d+6+4 n}{2}}
\exp \left(\frac{A}{g_{\rm r}} \right) \Bigg] \, .
\end{eqnarray}
The real parts are the familiar perturbative $\beta$-functions
\begin{eqnarray}
\beta_g^{\rm p}(g_{\rm r},v_{\rm r})\!\!&=&\!\!
-\varepsilon g_{\rm r}+\frac{1}{8\pi^2}
\left(-2 g_{\rm r} v_{\rm r}+v_{\rm r}^2+10 g_{\rm r}^2 \right)\, ,
\nonumber \\
\beta_v^{\rm p}(g_{\rm r},v_{\rm r})\!\!&=&\!\!
-\varepsilon v_{\rm r}+\frac{3}{8\pi^2} \left(
4g_{\rm r} v_{\rm r}-v_{\rm r}^2 \right)\, .
\end{eqnarray}
In the leading contributions the pair of
coupling constants $(g_{\rm r},v_{\rm r})$ may be replaced
by $(g_{\rm R},v_{\rm R})$ in (\ref{bet1}).
This can be proven using (\ref{REN}). The result is
\begin{eqnarray}
\label{betaR}
\beta_g (g_{\rm R},v_{\rm R})=
\beta_g^{\rm p}(g_{\rm R},v_{\rm R})
\!\!\!\!&-&\!\!\!\!i 2^7 3 \pi^4 c_{\rm r}
\displaystyle{\sum_{n=0}^{\infty}}
\Bigg[ \frac{v_{\rm R}^n}{n!} \left(\frac{3}{4\pi^2}\right)^n
B \left( n+\frac{1}{2},n+\frac{5}{2} \right)
\nonumber  \\
\nonumber \\
& &
\times \left(-\frac{A}{g_{\rm R}} \right)^{\frac{d+6+4 n}{2}}
\exp \left(\frac{A}{g_{\rm R}} \right) \Bigg]\, ,
\nonumber \\
\nonumber \\
\beta_v(g_{\rm R},v_{\rm R})=
\beta_v^{\rm p}(g_{\rm R},v_{\rm R})
\!\!\!\!&+&\!\!\!\!i 2^7 3 \pi^4 c_{\rm r}
\displaystyle{\sum_{n=0}^{\infty}}
\Bigg[ \frac{v_{\rm R}^n}{n!} \left(\frac{3}{4\pi^2}\right)^n
B \left(n+\frac{1}{2},n+\frac{5}{2} \right)
\nonumber  \\
\nonumber \\
& &
\times \left(\frac{4n}{2n+3}\right)
\left(-\frac{A}{g_{\rm R}} \right)^{\frac{d+6+4 n}{2}}
\exp \left(\frac{A}{g_{\rm R}} \right) \Bigg] \, .
\end{eqnarray}
Note that the $\beta$-functions have a form expected within a minimal
subtracted scheme:~${\beta}_g(g_{\rm R},v_{\rm R})=
-\varepsilon g_{\rm R} +{\beta}^{\, 4}_g(g_{\rm R},v_{\rm R})$
and ${\beta}_v(g_{\rm R},v_{\rm R})=
-\varepsilon v_{\rm R}+{\beta}^{\, 4}_v(g_{\rm R},v_{\rm R})$, where the
${\beta}^{\, 4}$ denotes the $\beta$-functions in four dimensions.
The critical
large-distance behavior of the correlation functions is defined by the
anomalous dimension of the field $\phi$, which is defined by
\begin{equation}
\gamma(g_{\rm R},v_{\rm R})=\mu \frac{\partial}{\partial \mu} \ln Z^{\phi}=
\mu \frac{\partial}{\partial \mu} \ln Z_{\rm p}^{\phi}+
\mu \frac{\partial}{\partial \mu} \ln Z_{{\rm np}}^{\phi} \, .
\end{equation}
With (\ref{zp}) for $Z^{\phi}_{\rm p}$ and (\ref{znp})
for $Z^{\phi}_{{\rm np}}$ we get the
$\varepsilon$-independent expression
\begin{eqnarray}
\label{gamR}
\lefteqn{\gamma(g_{\rm R},v_{\rm R}) =}
\nonumber \\
\nonumber \\
& &\!\!\!\!\!\frac{1}{2(8\pi^2)^2}
\left(4 g_{\rm R}^2-2 v_{\rm R} g_{\rm R} +v_{\rm R}^2 \right)
+i\,2^4 \pi^2 3 \,c_{\rm r}\,
\left(2^7 \pi^4 3^{-1}\,{\rm P}+\gamma-\frac{1}{2}-\ln2 \right)
\nonumber \\
\nonumber \\
& & \quad \times \sum_{n=0}^{\infty} \Bigg[\frac{v_{\rm R}^n}{n!}
\left(\frac{3}{4\pi^2}\right)^n B \left(n+\frac{1}{2},n+\frac{3}{2} \right)
\left(-\frac{A}{g_{\rm R}} \right)^{\frac{d+4+4n}{2}} \exp\left(
\frac{A}{g_{\rm R}}\right) \Bigg] \, .
\end{eqnarray}
The divergence of the length scale at critical temperature is governed by
the anomalous dimension of the composite field $\phi^2$:
\begin{equation}
\gamma_{\phi^2}(g_{\rm R},v_{\rm R})=
-\mu \frac{\partial}{\partial \mu} \ln Z^{\phi^2}=
-\mu \frac{\partial}{\partial \mu} \ln Z_{\rm p}^{\phi^2}
-\mu \frac{\partial}{\partial \mu} \ln Z_{{\rm np}}^{\phi^2} \, .
\end{equation}
Using (\ref{pPP}) and (\ref{npPP}), we obtain
\begin{eqnarray}
\label{Npp2}
\lefteqn{\gamma_{\phi^2}(g_{\rm R},v_{\rm R})=\frac{1}{8\pi^2}
\left(4 g_{\rm R}-v_{\rm R} \right)
+i\,2^4 \pi^2 3^2 \,c_{\rm r}\,
\left(\frac{8\pi^2}{3} {\rm Q}-\gamma+\ln2 \right)}
\nonumber \\
\nonumber \\
& & \quad \times \sum_{n=0}^{\infty} \Bigg[\frac{v_{\rm R}^n}{n!}
\left(\frac{3}{4\pi^2}\right)^n B \left(n+\frac{1}{2},n+\frac{3}{2} \right)
\left(-\frac{A}{g_{\rm R}} \right)^{\frac{d+6+4n}{2}} \exp\left(
\frac{A}{g_{\rm R}}\right) \Bigg].
\nonumber \\
\end{eqnarray}
For the leading contributions to the imaginary part, it was possible to replace
$(g_{\rm r},v_{\rm r})$ by $(g_{\rm R},v_{\rm R})$.
We remark that the expansions (\ref{betaR}),(\ref{gamR}) and (\ref{Npp2})
have the correct isotropic limit for $v_{\rm R} \rightarrow 0$.
\section{Generalization to the case $N=3$}
Generalizing (\ref{exp}) to $N=3$ we write:
\begin{eqnarray}
\vec{\phi}\!\!\!&=&\!\!\!\vec{u}_L \phi_c+\vec{u}_L \xi+\vec{u}_{T_1} \eta+
             \vec{u}_{T_2} \chi
\nonumber \\
\nonumber \\
       \!\!\!&=&\!\!\!
\left( \begin{array}{c}
       \sin{\theta} \cos{\vartheta} \\
       \sin{\theta} \sin{\vartheta} \\ \cos{\theta}
       \end{array} \right) (\phi_c+\xi)+
\left( \begin{array}{c}
       -\sin{\vartheta} \\ \cos{\vartheta} \\ 0
       \end{array} \right) \eta+
\left( \begin{array}{c}
       \cos{\theta} \cos{\vartheta} \\ \cos{\theta} \sin{\vartheta}
       \\ -\sin{\theta}
       \end{array} \right) \chi \, ,
\end{eqnarray}
where $\theta$ and $\vartheta$ are the rotation angles of the isotropic
instanton.
This change of variables yields the
following expansion of the energy functional:
\begin{eqnarray}
\label{Hex3}
H(\vec{\phi})=&H_c&\!\!\!+\frac{1}{2} \int d^dx \xi M_L \xi+
                \frac{1}{2} \int d^dx \eta M_{T_1} \eta+
                \frac{1}{2} \int d^dx \chi M_{T_2} \chi
                \nonumber \\
                \nonumber \\
              &-&\!\!\!\frac{\varepsilon 4(2)^{1/2}}{(-g)^{1/2}}
                \int d^dx\frac{\lambda^3}{[1+\lambda^2(x-x_0)^2]^2} \xi
                \nonumber \\
                \nonumber \\
              &-&\!\!\!
                (-8g)^{1/2} \int d^dx \frac{\lambda}{1+\lambda^2(x-x_0)^2}
                \left[\xi^3+(1-\delta)(\xi \eta^2+\xi \chi^2) \right]
                \nonumber \\
                \nonumber \\
              &+&\!\!\!
                \delta \,\, F(g,\vartheta,\theta,\xi,\eta,\chi)
\end{eqnarray}
with $M_L$ and $M_{T_1}=M_{T_2}=M_T$ of equation (\ref{MLT}).
For brevity, we have written down explicitly
only the terms responsible for the
leading contributions in the expansion (\ref{IMG}).
The remaining terms are denoted collectively by
$\delta \,\, F(g,\vartheta,\theta,\xi,\eta,\chi)$.
The classical contribution to the energy functional is
\begin{eqnarray}
H(\phi_{1c},\phi_{2c},\phi_{3c})\!\!\!&=&\!\!\!
-\frac{\lambda^{\varepsilon}}{g} \frac{8\pi^2}{3}
\left[1-\frac{\varepsilon}{2}\left(2+\ln \pi+\gamma \right) \right]
\nonumber \\
\nonumber \\
& &\!\!\!\!\!
-\frac{\lambda^{\varepsilon} \delta }{2g} \, \frac{8\pi^2}{3}
\left[\sin^4{\theta} \sin^2(2\vartheta)+\sin^2(2\theta) \right]+
{\cal O}\left(\frac{ \delta }{g}\, \varepsilon \right).
\end{eqnarray}
With similar steps as for $N=2$ we obtain
the bare imaginary parts of the vertex functions:
\begin{eqnarray}
\lefteqn{{\rm Im} \,\Gamma_{\rm b}^{(2M)} (q_m)_{i_1,i_2,\ldots,i_{2M}} = }
\nonumber \\
\nonumber \\
    & &\!\!\!\!\!
       \displaystyle{-c_{\rm b}^{(3)}} \int\limits_0^{\infty}
       \frac{d\lambda}{\lambda} \Bigg\{
       \lambda^{d-M(d-2)} \left(-\frac{\lambda^{\varepsilon}8\pi^2}{3g}
       \right)^{(d+3+2M)/2} \!\!\! \exp \left(\frac{\lambda^{\varepsilon}A}{g}
       \right) \prod^{2M}_{\nu=1}
       \left(\frac{q_{\nu}}{\lambda}\right)^2 \!
       \tilde{\phi} \left( \frac{q_{\nu}}{\lambda} \right)
       \nonumber  \\
       \nonumber \\
    & & \!\!\!\!\!
       \times \frac{1}{2} \int\limits_0^{2\pi} \int\limits_0^{\pi}
       \sin{\theta} d \theta d \vartheta \, \, u_{L\,i_1}(\vartheta,\theta)
       u_{L\,i_2}(\vartheta,\theta) \cdots u_{L\,i_{2M}} (\vartheta,\theta)
       \nonumber \\
       \nonumber \\
    & & \qquad \times
       \exp \left[a \sin^4{\theta} \sin^2(2\vartheta)+a \sin^2(2\theta) \right]
       \Bigg\} \left[1+{\cal O}(\varepsilon,g) \right] \, ,
\end{eqnarray}
where
\begin{eqnarray}
c_{\rm b}^{(3)}\!\!\!&=&\!\!\!
2^{-5/6} 3 \pi^{-13/3} \exp
                      \Bigg[
                       \frac{1}{3\varepsilon}
                       \left(11-4\delta+2{\delta}^2\right)+
                       \frac{5 \zeta^{'} (2)}{\pi^2}-
                       \frac{9}{2} \, \gamma -\frac{17}{4}
                       \nonumber  \\
                       \nonumber \\
         & &\qquad \qquad
            +\frac{16}{9} \, \delta -\frac{7}{81} \, {\delta}^2
            +{\cal O} \left( {\delta}^3 \right) \Bigg] \, ,
            \quad (i_k\, =\, 1,2,3)\, ,
\end{eqnarray}
with $a$, $A$ as in (\ref{aA}).
After evaluating the integrals over the collective coordinates
$\lambda$, $\theta$, and $\vartheta$ we obtain
for the imaginary part of the four-point vertex function
\begin{eqnarray}
\lefteqn{{\rm Im} \, \Gamma_{\rm b}^{(4)}(q_m)_{i,j,k,l}= } \nonumber \\
              \nonumber \\
&             &2^8 3^2 \pi^4 \,c_{\rm b}^{(3)}\,\, \frac{1}{\varepsilon}
              \sum_{n=0}^{\infty}
              \Bigg\{ \frac{(-2\, \delta)^n}{\Gamma(2n+7/2)}
              \left(\frac{g}{A}\right)
              \left(-\frac{\mu^{\varepsilon}A}{g}\right)^{\frac{d+7+2n}{2}}
              \!\!\! \exp \left(\frac{\mu^{\varepsilon}A}{g} \right)
              \nonumber \\
              \nonumber \\
&        &\!\!\!\times
          \sum_{p=0}^n \frac{\Gamma(3+2n-p)\,\Gamma\left(p+\frac{1}{2} \right)}
          {(n-p)!\, p! \, \Gamma\left(2(n-p)+3\right)}
          \Bigg[
          3\Gamma\left(n-p+\frac{3}{2}\right)^2 \,\,(S_{ijkl}-\delta_{ijkl})
          \nonumber \\
          \nonumber \\
&        & \qquad
         +\Gamma\left(n-p+\frac{1}{2}\right) \Gamma\left(n-p+\frac{5}{2}\right)
         \,\, \delta_{ijkl}
         \Bigg] \Bigg\}\, .
\end{eqnarray}
Similarly the bare imaginary parts of
$\frac{\partial}{\partial q^2}\Gamma^{(2)}(q)$
and $\Gamma^{(2,1)}(q_1,q_2;q_3)$ are found to be
\begin{eqnarray}
\lefteqn{\frac{\partial}{\partial q^2}
{\rm Im} \,\Gamma_{\rm b}^{(2)}(q)_{i,j}=
\delta_{ij} \, 2^4 \pi^2 3 \,c_{\rm b}^{(3)}
\left[\gamma-\frac{1}{2}+\ln \left(\frac{q}{2\mu}\right) \right]}
\nonumber \\
\nonumber \\
& & \times
\frac{1}{\varepsilon} \sum_{n=0}^{\infty} \Bigg[\frac{(-2\, \delta)^n}{n!}
 \, I(n) \,
\left(-\frac{\mu^{\varepsilon}A}{g} \right)^{\frac{d+3+2n}{2}} \exp\left(
\frac{\mu^{\varepsilon}A}{g}\right)
\Bigg] \, ,
\end{eqnarray}
and
\begin{eqnarray}
{\rm Im} \,\Gamma_{\rm b}^{(2,1)}(q_1,q_2;q_3)_{i,j}\!\!&=&\!\!
\delta_{ij} \, 2^4 \pi^2 3^2 \,c_{\rm b}^{(3)}\,
\frac{1}{\varepsilon} \sum_{n=0}^{\infty} \left\{
\frac{(-2\, \delta)^n}{n!} I(n)
\nonumber \right. \\
\nonumber \\
& &\left. \!\!\!\!\!\!\!\! \times
\left[\gamma+\ln \left(\frac{q_3}{2\mu}\right)\right]
\left(-\frac{\mu^{\varepsilon}A}{g} \right)^{\frac{d+5+2n}{2}} \exp\left(
\frac{\mu^{\varepsilon}A}{g}\right) \right\},
\nonumber \\
\end{eqnarray}
with
\begin{equation}
I(n) \equiv
\sum_{p=0}^n
\left( \begin{array}{c}
       n \\ p
       \end{array} \right)
B \left(2+2n-p,p+\frac{1}{2}\right)
B \left(n-p+\frac{1}{2},n-p+\frac{3}{2} \right).
\end{equation}
Now we apply the extended renormalization scheme. Using the coupling
constants $v$ and $g$, we obtain for the nonperturbative renormalized
couplings
\begin{eqnarray}
\label{c3R}
g_{\rm R}\!\!\!&=&\!\!\!g_{\rm r}
+i 2^7 3 \pi^4 c_{\rm r}^{(3)} \frac{1}{\varepsilon}
\displaystyle{\sum_{n=0}^{\infty}}
\Bigg[ \frac{v_{\rm r}^n}{n!} \left(\frac{3}{4\pi^2}\right)^n J_g(n)
\left(-\frac{A}{g_{\rm r}} \right)^{\frac{d+5+4 n}{2}}
\exp \left(\frac{A}{g_{\rm r}} \right) \Bigg] \, ,
\nonumber \\
\nonumber \\
v_{\rm R}\!\!\!&=&\!\!\!v_{\rm r}
-i 2^7 3 \pi^4 c_{\rm r}^{(3)} \frac{1}{\varepsilon}
\displaystyle{\sum_{n=0}^{\infty}}
\Bigg[ \frac{v_{\rm r}^n}{n!} \left(\frac{3}{4\pi^2}\right)^n J_v(n)
\left(-\frac{A}{g_{\rm r}} \right)^{\frac{d+5+4 n}{2}}
\exp \left(\frac{A}{g_{\rm r}} \right) \Bigg]\, ,
\nonumber \\
\end{eqnarray}
with
\begin{eqnarray}
\label{J}
\displaystyle{ J_g(n)}
\!\!\!&\equiv&\!\!\!
\sum_{p=0}^n
\left( \begin{array}{c}
       n \\ p
       \end{array} \right)
B \left(3+2n-p,p+\frac{1}{2}\right)
B \left(n-p+\frac{1}{2},n-p+\frac{5}{2} \right)\, ,
\nonumber \\
\nonumber \\
\displaystyle{ J_v(n)}
\!\!\!&\equiv&\!\!\!
\sum_{p=0}^n \Bigg\{
\left( \begin{array}{c}
       n \\ p
       \end{array} \right)
\, \left[ \frac{4(n-p)}{2(n-p)+3} \right] \,
B \left(3+2n-p,p+\frac{1}{2}\right)
\nonumber \\
\nonumber \\
& & \qquad \qquad \qquad \qquad \quad \times
B \left(n-p+\frac{1}{2},n-p+\frac{5}{2} \right) \Bigg\} \, ,
\end{eqnarray}
and
\begin{equation}
\label{cc33}
c_{\rm r}^{(3)}=3\,\, 2^{-5/6} \pi^{-5/2} \exp \bigg[
\frac{5\zeta^{'}(2)}{\pi^2}-\frac{8}{3} \, \gamma-\frac{7}{12}
+{\cal O}({\delta}_{\rm r})+{\cal O}({\delta}_{\rm r} \, g_{\rm r}) \bigg]\, .
\end{equation}
For the nonperturbative renormalization constants $Z^{\phi}$ and $Z^{\phi^2}$
we find
\begin{eqnarray}
Z_{{\rm np}}^{\phi}\!\!&=&\!\!1-
i\,2^4 \pi^2 3 \,c_{\rm r}^{(3)}\,
\frac{1}{\varepsilon} \left(2^4 \pi^2 A\,\,{\rm P}+\gamma-\frac{1}{2}-\ln2
\right)
\nonumber \\
\nonumber \\
& & \mbox{} \times \sum_{n=0}^{\infty} \Bigg[\frac{v_{\rm R}^n}{n!}
\left(\frac{3}{4\pi^2}\right)^n \, I(n) \,
\left(-\frac{A}{g_{\rm R}} \right)^{\frac{d+3+4n}{2}} \exp\left(
\frac{A}{g_{\rm R}}\right) \Bigg] \,
\end{eqnarray}
and
\begin{eqnarray}
\lefteqn{Z_{{\rm np}}^{\phi} Z_{{\rm np}}^{\phi^2}=1+
i\,2^4 \pi^2 3^2 \,c_{\rm r}^{(3)}\,
\frac{1}{\varepsilon} \left(\frac{8\pi^2}{3} {\rm Q}-\gamma+\ln2 \right)}
\nonumber \\
\nonumber \\
& &\times \sum_{n=0}^{\infty} \Bigg[\frac{v_{\rm R}^n}{n!}
\left(\frac{3}{4\pi^2}\right)^n I(n)
\left(-\frac{A}{g_{\rm R}} \right)^{\frac{d+5+4n}{2}} \exp\left(
\frac{A}{g_{\rm R}}\right) \Bigg].
\end{eqnarray}
To calculate the nonperturbative renormalization
group functions $\beta(g_{\rm R},v_{\rm R})$, $\gamma(g_{\rm R},v_{\rm R})$
and $\gamma_{\phi^2}(g_{\rm R},v_{\rm R})$,
we simply repeat the calculations of the case $N=2$.
Using the expressions (\ref{c3R}) for $g_{\rm R}$ and $v_{\rm R}$ yields the
$\beta$-functions
\begin{eqnarray}
\label{B3}
\lefteqn{\beta_g (g_{\rm R},v_{\rm R})=}
\nonumber \\
& &\beta_g^{\rm p}(g_{\rm R},v_{\rm R})
-i 2^7 3 \pi^4 c_{\rm r}^{(3)}
\displaystyle{\sum_{n=0}^{\infty}}
\Bigg[ \frac{v_{\rm R}^n}{n!} \left(\frac{3}{4\pi^2}\right)^n J_g(n)
\left(-\frac{A}{g_{\rm R}} \right)^{\frac{d+7+4 n}{2}}
\exp \left(\frac{A}{g_{\rm R}} \right) \Bigg]\, ,
\nonumber \\
\nonumber \\
\lefteqn{\beta_v(g_{\rm R},v_{\rm R})=}
\nonumber \\
& &\beta_v^{\rm p}(g_{\rm R},v_{\rm R})
+i 2^7 3 \pi^4 c_{\rm r}^{(3)}
\displaystyle{\sum_{n=0}^{\infty}}
\Bigg[ \frac{v_{\rm R}^n}{n!} \left(\frac{3}{4\pi^2}\right)^n J_v(n)
\left(-\frac{A}{g_{\rm R}} \right)^{\frac{d+7+4 n}{2}}
\exp \left(\frac{A}{g_{\rm R}} \right) \Bigg]\, .
\nonumber \\
\nonumber \\
\end{eqnarray}
The expressions for the nonperturbative anomalous dimension
of the field $\phi$ and the composite field $\phi^2$ are given by
\begin{eqnarray}
\label{G3}
\gamma(g_{\rm R},v_{\rm R})\!\!\!\!&=&\!\!\!\!
\gamma^{\rm p}(g_{\rm R},v_{\rm R})
+i\,2^4 \pi^2 3 \,c_{\rm r}^{(3)}\,
\left(2^7 \pi^4 3^{-1}\,{\rm P}+\gamma-\frac{1}{2}-\ln2 \right)
\nonumber \\
\nonumber \\
& & \quad \times
\sum_{n=0}^{\infty} \Bigg[\frac{v_{\rm R}^n}{n!}
\left(\frac{3}{4\pi^2}\right)^n I(n)
\left(-\frac{A}{g_{\rm R}} \right)^{\frac{d+5+4n}{2}} \exp\left(
\frac{A}{g_{\rm R}}\right) \Bigg] \, ,
\end{eqnarray}
and
\begin{eqnarray}
\label{Npp3}
\gamma_{\phi^2}(g_{\rm R},v_{\rm R})\!\!&=&\!\!
\gamma_{\phi^2}^{\rm p}(g_{\rm R},v_{\rm R})
+i\,2^4 \pi^2 3^2 \,c_{\rm r}^{(3)}\,
\left(\frac{8\pi^2}{3} {\rm Q}-\gamma+\ln2 \right)
\nonumber \\
\nonumber \\
\!\!&\times&\!\! \sum_{n=0}^{\infty} \Bigg[\frac{v_{\rm R}^n}{n!}
\left(\frac{3}{4\pi^2}\right)^n I(n)
\left(-\frac{A}{g_{\rm R}} \right)^{\frac{d+7+4n}{2}} \exp\left(
\frac{A}{g_{\rm R}}\right) \Bigg] \, ,
\end{eqnarray}
respectively.
\section{Discussion and Conclusions}
Our main results are the real and imaginary parts (\ref{B3})--(\ref{Npp3})
of the renormalization-group functions $\beta$, $\gamma$ and $\gamma_{\phi^2}$
in terms of the nonperturbative renormalized
couplings $g_{\rm R}$ and $v_{\rm R}$. Via the dispersion relation (\ref{DR}),
we obtain the large-order behavior (\ref{LO2}).
It is useful to check the expansion terms proportional to powers
of $1/|g|$ accompanying the $v^n$ in the imaginary part of
each renormalization-group function.
If the power of $1/|g|$ is denoted by $p(n)$, we note that $p(n)$
is the same as for the corresponding vertex function
before the integration over the dilatation parameter $\lambda$.
This integration with the measure $d\ln \lambda$
reduces $p(n)$ by $1$ [see (\ref{lambda})].
However this effect is compensated by the differentiation with respect
to the logarithm of the scale parameter $\mu$ in the definition of the
renormalization group functions.
The power $p(n)$ in the vertex functions can simply be explained:
For a $2M$-point vertex function with $k$ $\phi^2$-insertions
there is first a power $(1/|g|)^{(M+k)}$ from the fields.
This is multiplied by a factor $(1/|g|)^{(d+(N-1)+1)/2}$ from the $d$
translational, $(N-1)$ rotational, and one dilatational would-be zero modes.
The classical contribution to the energy functional yields
a factor $(1/|g|)^2$ for each power of the anisotropic constant $v$.
Hence, for the renormalization
group function which follows at the one-loop level from the vertex
function $\Gamma^{(2M,k)}$, the value of $p(n)$ is given by
\begin{equation}
p(n)=\frac{d+N+2(M+k)+4n}{2} \, .
\end{equation}

Inserting the imaginary parts into a dispersion relation (\ref{DR}),
we can estimate the large-order coefficients of the corresponding
series expansions
\begin{equation}
f(g,v)=\sum_{k,n=0}^{\infty} f_{kn}\, g^k \, v^n \, ,
\end{equation}
where $n \ll k$.
The function $f$ stands for
${\beta}_g$, ${\beta}_v$, $\gamma$ and $\gamma_{\phi^2}$, respectively.
The rules how to go from the imaginary parts of $f$ to $f_{kn}$ are
known from (\ref{LO2}).
For instance, the large-order coefficients of the $\beta$-functions for $N=3$
are given by
\begin{eqnarray}
\label{concl}
\beta_{g, \,kn}\!\!&=&\!\!c_{\rm r}^{(3)}2^7 3 \pi^3
\left(\frac{3}{4\pi^2}\right)^n
\frac{ J_g(n) }{n!}
\nonumber \\
\nonumber \\
& &\qquad \times
(-1)^k \left(\frac{3}{8\pi^2}\right)^k k! \, k^{2n+(d+5)/2}
\left[1+{\cal O}(1/k)\right] \, ,
\nonumber \\
\nonumber \\
\beta_{v, \,kn}\!\!&=&\!\!-c_{\rm r}^{(3)}2^7 3 \pi^3
\left(\frac{3}{4\pi^2}\right)^n
\frac{ J_v(n) }{n!}
\nonumber \\
\nonumber \\
& &\qquad \times
(-1)^k \left(\frac{3}{8\pi^2}\right)^k k! \, k^{2n+(d+5)/2}
\left[1+{\cal O}(1/k)\right] \, ,
\end{eqnarray}
with $J$, $c_{\rm r}^{(3)}$ from (\ref{J}) and (\ref{cc33}), respectively.

The leading large-$k$ behavior can be checked by means of a simple
combinatorial analysis. We start with a series in terms of the standard
couplings $u$ and $v$ [see (\ref{HN})]. For illustration,
we consider the contribution of the pure $u$-powers.
It is known from a theory with only one coupling constant, that the
coefficients have the large-order behavior
\begin{equation}
\label{lao}
f_l \, u^l \longrightarrow \gamma \, (-\alpha)^l \,
\Gamma(l+b+1)\, u^l \qquad (l \gg 1) \, .
\end{equation}
Changing over to our couplings ($u = g-v$),
the right-hand side of (\ref{lao}) contributes
\begin{equation}
\gamma \, (-1)^{l-k}\, (-\alpha)^l \, \Gamma(l+b+1) \,
\left(\begin{array}{c}
       l \\ k
      \end{array} \right) \, g^k \, v^{l-k} \, ,
\end{equation}
with $$\left(\begin{array}{c}
        l \\ k
       \end{array} \right)=\frac{\Gamma(l+1)}{k!\, (l-k)!} \, .$$
Replacing $l$ by $l=n+k$, the contribution of $g^k v^n$ has the form
\begin{equation}
f_{kn}\, g^k \, v^n \longrightarrow \gamma \, (\alpha)^n \, (-\alpha)^k \,
\frac{\Gamma(k+n+b+1)\, \Gamma(k+n+1)}{k! \, n!} \, g^k \, v^n \, .
\end{equation}
Now, for large $k$, the $\Gamma$-functions can be approximated by
\begin{equation}
\Gamma(k+\delta+1) \longrightarrow
k! \, k^{\delta} \, \left(1+{\cal O}(1/k) \right) \, .
\end{equation}
Thus in the region $n \ll k$ we obtain
\begin{equation}
f_{kn} \sim (-\alpha)^k \, k! \, k^{2n+b} \, ,
\end{equation}
in agreement with (\ref{concl}).

As stated in the introduction, the stable cubic fixed point is expected to lie
in the vicinity of the isotropic fixed point.
Since $v$ becomes very small in this region, reasonable results should
be obtained by resumming the $g$-series accompanying each power $v^n$.
This will be done in a forthcoming publication.

\end{document}